\documentclass[useAMS,usenatbib]{mn2e}

\usepackage{times}
\usepackage{amssymb}
\usepackage{caption}
\usepackage{lscape}
\usepackage[fleqn]{amsmath}
\usepackage{url}
\usepackage{graphicx}
\usepackage{subfigure}
\usepackage{float}
\usepackage{xcolor}
\usepackage{bm}
\usepackage{enumitem}
\usepackage[figuresright]{rotating}
\usepackage{txfonts}
\usepackage{xspace}

\newcommand{\SimpleX}{{\sc SimpleX}\xspace}
\newcommand{\MdotA}{\ifmmode{\dot{M}_{\eta_{\rm A}}}\else{$\dot{M}_{\eta_{\rm A}}$}\fi\xspace}
\newcommand{\MdotB}{\ifmmode{\dot{M}_{\eta_{\rm B}}}\else{$\dot{M}_{\eta_{\rm B}}$}\fi\xspace}
\newcommand{\etaA}{$\eta_{\rm A}$\xspace}
\newcommand{\etaB}{$\eta_{\rm B}$\xspace}
\newcommand{\ec}{$\eta$~Car\xspace}

\newcommand{\Ms}{\ifmmode{~{\rm M}_{\odot}}\else{${\rm M}_{\odot}$}\fi\xspace}
\newcommand{\Msy}{\ifmmode{\Ms\,{\rm yr}^{-1}}\else {$\Ms\,{\rm yr}^{-1}$}\fi\xspace}
\newcommand{\Ls}{\ifmmode{~{\rm L}_{\odot}}\else{${\rm L}_{\odot}$}\fi\xspace}
\newcommand{\kms}{\ifmmode{~{\rm km\,s}^{-1}}\else {${\rm km\,s}^{-1}$}\fi\xspace}
\newcommand{\Rs}{\ifmmode{~{\rm R}_{\odot}}\else{${\rm R}_{\odot}$}\fi\xspace}

\newcommand{\HI}{\ifmmode{{\rm H\,I}}\else{H{\sc$\,$i}}\fi\xspace}
\newcommand{\HII}{\ifmmode{{\rm H\,II}}\else{H{\sc$\,$ii}}\fi\xspace}
\newcommand{\HeI}{\ifmmode{{\rm He\,I}}\else{He{\sc$\,$i}}\fi\xspace}
\newcommand{\HeII}{\ifmmode{{\rm He\,II}}\else{He{\sc$\,$ii}}\fi\xspace}
\newcommand{\HeIII}{\ifmmode{{\rm He\,III}}\else{He{\sc$\,$iii}}\fi\xspace}
\newcommand{\FeIII}{\ifmmode{[{\rm Fe\,III}]}\else{[Fe{\sc$\,$iii}]}\fi\xspace}
\newcommand{\FeII}{\ifmmode{[{\rm Fe\,II}]}\else{[Fe{\sc$\,$ii}]}\fi\xspace}
\newcommand{\NiII}{\ifmmode{[{\rm Ni\,II}]}\else{[Ni{\sc$\,$ii}]}\fi\xspace}

\defcitealias{Madura_etA_2012}{M12}
\defcitealias{Madura_etA_2013}{M13}

\title[3D Radiative Transfer in $\eta$ Carinae]{3D Radiative Transfer in $\eta$~Carinae: Application of the SimpleX Algorithm to 3D SPH Simulations of Binary Colliding Winds}

\author[Clementel et al.]{N. Clementel$^{1}$\thanks{E-mail: clementel@strw.leidenuniv.nl}, T.~I. Madura$^{2}$,  C.~J.~H. Kruip$^{1}$, V. Icke$^{1}$, T.~R. Gull$^{2}$\\
$^{1}$Leiden Observatory, Leiden University, P.O. Box 9513, 2300 RA Leiden, the Netherlands\\
$^{2}$Astrophysics Science Division, Code 667, NASA Goddard Space Flight Center, Greenbelt, MD 20771, USA\\}

\begin{document}

\date{Accepted 2014 June 24. Received 2014 June 20; in original form 2014 February 23}

\pagerange{\pageref{firstpage}--\pageref{lastpage}} \pubyear{2014}

\maketitle

\label{firstpage}

\begin{abstract}
Eta Carinae is an ideal astrophysical laboratory for studying massive binary interactions and evolution, and stellar wind-wind collisions.
Recent three-dimensional (3D) simulations set the stage for understanding the highly complex 3D flows in \ec .
Observations of different broad high- and low-ionization forbidden emission lines provide an excellent tool to constrain the orientation of the system, the primary's mass-loss rate, and the ionizing flux of the hot secondary. In this work we present the first steps towards generating synthetic observations to compare with available and future \emph{HST}/STIS data.
We present initial results from full 3D radiative transfer simulations of the interacting winds in \ec. We use the \SimpleX algorithm to post-process the output from 3D SPH simulations and obtain the ionization fractions of hydrogen and helium assuming three different mass-loss rates for the primary star. The  resultant ionization maps of both species constrain the regions where the observed forbidden emission lines can form.
Including collisional ionization is necessary to achieve a better description of the ionization states, especially in the areas shielded from the secondary's radiation. We find that reducing the primary's mass-loss rate increases the volume of ionized gas, creating larger areas where the forbidden emission lines can form. We conclude that post processing 3D SPH data with \SimpleX is a viable tool to create ionization maps for \ec.
\end{abstract}



\section{Introduction}

Eta Carinae (\ec) is an extremely luminous ($L_{\rm Total} \gtrsim 5 \times 10^6 \Ls$) colliding wind binary with a highly eccentric (e~$\sim 0.9$), 5.54 year orbit \citep{DavidsonHumphreys_1997, Damineli_etA_1997, Hillier_etA_2001, Damineli_etA_2008_b, Damineli_etA_2008_a, Corcoran_etA_2010}. \etaA, the primary of the system, is our closest ($2.3 \pm 0.1$ kpc, \citealt{Smith_2006}) example of a very massive star ($\sim100 \Ms$, \citealt{DavidsonHumphreys_1997}). A Luminous Blue Variable (LBV), \etaA\ has an extremely powerful stellar wind with $\MdotA \approx 8.5 \times 10^{-4} \Msy$ and $v_{\infty} \approx 420 \kms$ \citep{Hillier_etA_2001, Hillier_etA_2006, Groh_etA_2012}. Observations over the last two decades indicate that \etaA's dense stellar wind interacts with the hotter, less luminous companion star \etaB and its much faster ($v_{\infty} \approx 3000 \kms$, \citealt{Pittard_etA_2002}), but much lower density ($\MdotB \approx 10^{-5} \Msy$), wind \citep{Damineli_etA_2008_b, Corcoran_etA_2010, Gull_etA_2009, Gull_etA_2011}. These wind-wind interactions lead to various forms of time-variable emission and absorption seen across a wide range of wavelengths \citep{Damineli_etA_2008_b}.

Observational signatures that arise as a result of the wind-wind interactions are important for studying \ec as they provide crucial information about the physical properties of the as-yet unseen \etaB and the system as a whole. Three-dimensional (3D) hydrodynamical simulations show that the fast wind of \etaB carves a low-density cavity out of the slower, denser inner wind of \etaA for most of the orbit \citep{Okazaki_etA_2008, Parkin_etA_2011, Madura_Groh_2012, Madura_etA_2012, Madura_etA_2013, Russell_2013}. The same simulations indicate that the hot post-shock gas in the inner wind-wind interaction region (WWIR) gives rise to hard (up to 10~keV) X-ray emission that varies over the 5.54-year period. Together with the models, spatially unresolved X-ray \citep{Hamaguchi_etA_2007, Henley_etA_2008, Corcoran_etA_2010}, optical \citep{Damineli_etA_2008_b, Damineli_etA_2008_a}, and near-infrared \citep{Whitelock_etA_2004, Groh_etA_2010} observations have helped constrain the geometry and physical conditions within the inner WWIR.

In addition to the `current' interaction between the two winds that occurs in the inner regions (at spatial scales comparable to the semi-major axis length $a \approx 15.4$~AU $\approx 0.0067$~arcsec at 2.3~kpc), larger scale ($\approx 3250$~AU $\approx 1.4$~arcsec in diameter) 3D hydrodynamical simulations exhibit outer WWIRs that extend thousands of AU from the central stars \citep[][hereafter M12 and M13, respectively]{Madura_etA_2012, Madura_etA_2013}. Long-slit spectral observations of \ec with the \emph{Hubble Space Telescope}/Space Telescope Imaging Spectrograph (\emph{HST}/STIS) reveal these spatially-extended WWIRs, seen via emission from multiple low- and high-ionization forbidden lines \citep{Gull_etA_2009, Gull_etA_2011, Teodoro_etA_2013}.

Using a 3D dynamical model of the broad, extended \FeIII emission observed in \ec by the \emph{HST}/STIS, \citetalias{Madura_etA_2012} confirmed the orbital inclination and argument of periapsis that \citet{Okazaki_etA_2008} and \citet{Parkin_etA_2009} derived using X-ray data. More importantly, \citetalias{Madura_etA_2012} broke the degeneracy inherent to models based solely on X-rays or other spatially-unresolved data and constrained, for the first time, the 3D orientation of \ec's binary orbit.
\citetalias{Madura_etA_2012} find that the system has an argument of periapsis $\omega \approx 240^{\circ}$ to $285^{\circ}$, with the orbital axis closely aligned with the Homunculus nebula's polar axis at an inclination $i \approx 130^{\circ}$ to $145^{\circ}$ and position angle on the sky $\mathrm{PA} \approx 302^{\circ}$ to $327^{\circ}$, implying that apastron is on the observer's side of the system and that \etaB orbits clockwise on the sky. The dynamical model of \citetalias{Madura_etA_2012} was based on 3D smoothed particle hydrodynamics (SPH) simulations of \ec's colliding winds. A simple radiative transfer (RT) code was used to integrate the optically thin \FeIII emission and generate synthetic slit-spectra for comparison to the available \emph{HST}/STIS data. Although very successful, \citetalias{Madura_etA_2012} used a semi-analytic approach to compute the volume of wind material photoionized by \etaB. Furthermore, the fraction of Fe$^{2+}$ as a function of $T$ was estimated assuming collisional ionization equilibrium and available ion fraction data. Due to the lack, at the time, of a suitable code, proper 3D RT simulations of \etaB's ionizing radiation were not performed. The location and strength of the \FeIII emission was thus based on geometrical criteria, while in reality, the population of forbidden states depends on the local ionization state of the medium.

The goal of this paper is to improve considerably the modeling approach of \citetalias{Madura_etA_2012} by computing full 3D RT simulations of the effects of \etaB's ionizing radiation on \ec's spatially-extended WWIRs. We accomplish this by applying the \SimpleX algorithm for 3D RT on an unstructured Delaunay grid \citep{Ritzerveld_etA_2006, Ritzerveld_2007, Paardekooper_etA_2010, Paardekooper_etA_2011, Kruip_etA_2010} to recent 3D SPH simulations of \ec's binary colliding winds \citepalias{Madura_etA_2013}. We use \SimpleX to obtain detailed ionization fractions of hydrogen and helium at the resolution of the original SPH simulations. This should allow us to predict much more precisely where, and to what extent, various observed forbidden emission lines form. This paper lays the foundation for future work aimed at generating synthetic spectral data cubes for comparison to data obtained with \emph{HST}/STIS as part of a multi-cycle program to map changes in \ec's extended wind structures across one binary cycle from 2009 through 2015 \citep{Gull_etA_2011, Teodoro_etA_2013}. Comparison of the observations to the models should ultimately lead to more accurate constraints on the orbital, stellar, and wind parameters of the \ec system, such as \etaA's mass-loss rate and \etaB's temperature and luminosity (\citealt{Mehner_etA_2010, Mehner_etA_2012}; \citetalias{Madura_etA_2013}).

While we focus specifically on the case of \ec, the numerical methods in this paper can be applied to numerous other colliding wind (e.g. WR~140, WR~137, WR~19, \citealt{Fahed_etA_2011, Lefevre_etA_2005, Williams_etA_2009}) and dusty `pinwheel' (WR~104, WR~98a, \citealt{Tuthill_etA_1999, Monnier_etA_1999}) binary systems. One of the biggest remaining mysteries is how dust can form and survive in such systems that contain a hot, luminous O star. Coupled with 3D hydrodynamical simulations, \SimpleX simulations have the potential to help determine the regions where dust can form and survive in these unique objects.

In the following section we describe our numerical approach, including the SPH simulations, the \SimpleX code, and the RT simulations. Section~\ref{sec:Results} describes the results. A discussion of the results and their implications follows in Section~\ref{sec:Discussion}. Section~\ref{sec:Summary} summarizes our conclusions and outlines the direction for future work.


\section{Codes and Simulations}

\subsection{The 3D SPH Simulations}\label{ssec:SPH}

The hydrodynamical simulations were performed with the same SPH code used in \citetalias{Madura_etA_2013}, to which we refer the reader for details. Optically thin radiative cooling is implemented using the Exact Integration scheme of \citet{Townsend_2009}, with the radiative cooling function $\Lambda(T)$ calculated using {\sc Cloudy} 90.01 \citep{Ferland_etA_1998} for an optically thin plasma with solar abundances. The pre-shock stellar winds and rapidly-cooling dense gas in the WWIR are assumed to be maintained at a floor temperature $= 10^4$~K due to photoionization heating by the stars \citep{Parkin_etA_2011}. The same initial wind temperature ($T_{\rm wind}$) is assumed for both stars. The effect of $T_{\rm wind}$ on the flow dynamics is negligible \citep{Okazaki_etA_2008}.

Radiative forces are incorporated in the SPH code via an `anti-gravity' formalism, the details of which can be found in \citetalias{Madura_etA_2013} and \citet{Russell_2013}. The individual stellar winds are parametrized using the standard `beta-velocity law' $v(r) = v_{\infty}(1-R_{\star}/r)^{\beta}$, where $v_{\infty}$ is the wind terminal velocity, $R_{\star}$ the stellar radius, and $\beta$ (set $=1$) a free parameter describing the steepness of the velocity law. Effects due to `radiative braking' \citep{Gayley_etA_1997, Parkin_etA_2011}, photospheric reflection \citep{Owocki_2007}, and self-regulated shocks \citep[in which ionizing X-rays from the WWIR inhibit the wind acceleration of one or both stars, leading to lower pre-shock velocities and lower shocked plasma temperatures,][]{Parkin_etA_2013}, are not included. These effects are not expected to play a prominent role in \ec at the orbital phases near apastron considered in this work (\citealt{Parkin_etA_2009, Parkin_etA_2011, Russell_2013}; \citetalias{Madura_etA_2013}).We include the more important velocity-altering effects of `radiative inhibition', in which one star's radiation field reduces the net rate of acceleration of the opposing star's wind \citep{Stevens_etA_1994, Parkin_etA_2009, Parkin_etA_2011}. However, because we fix the mass-loss rates in our anti-gravity approach, possible changes to the mass-loss due to radiative inhibition are not included. These changes are not expected to be significant in \ec and should not greatly affect our results or conclusions \citepalias{Madura_etA_2013}.

Using an $xyz$ Cartesian coordinate system, the binary orbit is set in the $xy$ plane, with the origin at the system centre-of-mass and the major axis along the $x$-axis. The two stars orbit counter-clockwise when viewed from the $+z$-axis. By convention, $t = 0$ ($\phi = t/2024 = 0$) is defined as periastron. Simulations are started at apastron and run for multiple consecutive orbits. Orbits are numbered such that $\phi = 1.5$, 2.5 and 3.5 correspond to apastron at the end of the second, third, and fourth full orbits, respectively.

The outer spherical simulation boundary is set at $r = 100\,a$ from the system centre-of-mass, where $a = 15.45$~AU is the length of the orbital semimajor axis. Particles crossing this boundary are removed from the simulations. The computational domain is comparable in size to past and planned \emph{HST}/STIS mapping observations of the interacting stellar winds in \ec 's central core ($\sim \pm 0.67'' \simeq \pm1540$~AU, \citealt{Gull_etA_2011}; \citetalias{Madura_etA_2012}; \citealt{Teodoro_etA_2013}). As demonstrated by \citet{Gull_etA_2011} and \citetalias{Madura_etA_2012}, 3D simulations at this scale are necessary for understanding and modelling the extended, time-variable forbidden line emission structures that are spatially and spectrally resolved by \emph{HST}/STIS.

\begin{table}
\caption{Stellar, wind, and orbital parameters of the 3D SPH simulations}
\label{tab:etaParams}
\begin{center}
\begin{tabular}{l c c c}\hline
  Parameter & \etaA & \etaB & Reference \\ \hline
  $M_{\star}$ (\Ms) & 90 & 30 & H01; O08 \\
  $R_{\star}$ (\Rs) & 60 & 30 & H01; H06 \\
  $T_{\rm wind}$ ($10^4$ K) & 3.5 & 3.5 & O08; M13 \\
  $\dot{M}$ ($10^{-4} \Msy$) & 8.5, 4.8, 2.4 & 0.14 & G12; P09\\
  $v_{\infty}$ (\kms) & 420 & 3000 & G12; P02\\
  $P_{\rm orb}$ (days) & \multicolumn{2}{c}{2024} & D08a\\
  $e$ & \multicolumn{2}{c}{0.9} & C01; P09\\
  $a$ (AU) & \multicolumn{2}{c}{15.45} & C01; O08 \\\hline
\end{tabular}
\end{center}
\textbf{Notes:} $M_{\star}$ and $R_{\star}$ are the stellar mass and radius. $T_{\rm wind}$ is the initial wind temperature. $\dot{M}$ and $v_{\infty}$ are the stellar-wind mass-loss rate and terminal speed, respectively. $P_{\rm orb}$ is the period, $e$ is the eccentricity,  and $a$ is the length of the orbital semimajor axis.\\
\textbf{References:} C01 = \citet{Corcoran_etA_2001}; H01 = \citet{Hillier_etA_2001}; P02 = \citet{Pittard_etA_2002}; H06 = \citet{Hillier_etA_2006}; D08a = \citet{Damineli_etA_2008_a}; O08 = \citet{Okazaki_etA_2008}; P09 = \citet{Parkin_etA_2009}; G12 = \citet{Groh_etA_2012}.
\end{table}

The total number of SPH particles used in the simulations is roughly between $5 \times 10^5$ and $9 \times 10^5$, depending on the value of \MdotA. The adopted simulation parameters (Table \ref{tab:etaParams}) are consistent with those derived from the available observations, although there has been some debate on the exact present-day value of \MdotA (see \citetalias{Madura_etA_2013} for details). In an effort to better constrain \etaA's current $\dot{M}$, \citetalias{Madura_etA_2013} performed multiple 3D SPH simulations assuming different \MdotA. We use the same naming convention as \citetalias{Madura_etA_2013} when referring to the simulations in this paper for the different \MdotA, namely, Case~A ($\MdotA = 8.5 \times 10^{-4}$~\Msy), Case~B ($\MdotA = 4.8 \times 10^{-4}$~\Msy), and Case~C ($\MdotA = 2.4 \times 10^{-4}$~\Msy). We discuss the effects of the three values of \MdotA on the RT calculations in Section~\ref{ssec:Ion&Mdot}.


\subsection{The \SimpleX Algorithm for Radiative Transfer on an Unstructured Mesh}\label{ssec:SimpleX}

The \SimpleX algorithm, conceived by \citet{Ritzerveld_etA_2006}, implemented by \citet{Ritzerveld_2007}, and further improved by \citet{Paardekooper_etA_2010} and  \citet{Kruip_etA_2010}, is designed to solve the general equations of particle transport by expressing them as a walk on a graph. At the basis of the method lies the unstructured grid on which the photons are transported. A given medium (e.g. a density or optical-depth field) is typically sampled with a Poisson point process and the resulting point distribution is used to tessellate space according to the Voronoi recipe: all points in a cell are closer to the nucleus of that cell than to any other nucleus. The Voronoi nuclei are then connected by a Delaunay triangulation. The grid is constructed to describe the properties of the underlying physical medium, through which the photons travel, in such a way that more grid points are placed in regions with a higher opacity. The result is a higher resolution in places where it is needed most, i.e. where the optical depth is highest.

Photons are transported from node to node along the edges of the Delaunay triangulation, where each transition has a given probability. In one computational cycle, every nucleus in the grid transports its content to neighbouring nuclei, optionally absorbing or adding photons. Which neighbours are selected for transport depends on the specific process.
Even though \SimpleX was originally developed for application in cosmological radiative transfer, its properties are still well suited for our purpose.

In Section~\ref{ssec:Grid} we present the construction procedure for the \SimpleX RT mesh starting from the SPH particle distribution. Section~\ref{ssec:Chemistry} describes the processes that determine the ionization state of the gas, such as collisional- and photo-ionization and recombination, plus the specifics of their implementation in \SimpleX (for further details see Chapters 4 and 5 of \citealt{Kruip_2011}).


\subsubsection{Grid Construction}\label{ssec:Grid}

The 3D SPH simulations provide the time-dependent 3D density, temperature, and velocity structure of \ec 's interacting winds on the spatial scales of interest, thus forming the basis of our model. As in \citetalias{Madura_etA_2012}, the radiative transfer calculations are performed as post-processing of the 3D SPH output. The hydrodynamic influence of the radiation on the gas is thereby neglected. We do not expect this to have a very large influence on the results since the material photoionized by \etaB responds nearly instantaneously to its UV flux, i.e. the recombination time-scale is very small relative to the orbital time-scale, especially around apastron (M12).

The first step is to convert the SPH particle distribution to a \SimpleX mesh. Since the density field is given by discrete particles, we might obtain an estimate of the density at any position in the domain using a typical SPH kernel function $W(r, h)$ with

\begin{equation}\label{eq:kernel}
\rho(r)=\sum_j{m_jW(|r-r_j|,h_i)}
\end{equation}\\
where $h$ is the smoothing length and $m_j$ is the mass of particle $j$. Using this kernel function one can then sample the data using the sampling functions described in e.g. \citet{Paardekooper_etA_2010} and \citet{Kruip_etA_2010}.
However, given the fact that the original data is already particle-based, it is more natural for us to use the SPH particles themselves as the generating nuclei for the Voronoi-Delaunay mesh. This leads to a more direct estimate of the density, given by the division of the particle mass by the Voronoi volume of its corresponding cell. Another advantage is that, due to pressure forces, the particles in an SPH simulation are in general positioned more regularly than for a pure Poisson process. Finally, we note that with future applications of 3D time-dependent radiation-hydrodynamics in mind, a coupling of \SimpleX with an SPH method is most natural when the radiation transport is applied directly to the SPH particles so that no spurious interpolation is needed. For these reasons we use every SPH particle as the nucleus of a Voronoi cell. This procedure yields density estimates that are less smooth than those obtained with typical kernel functions of the type of Equation (\ref{eq:kernel}), but guarantees mass conservation and represents small scale structures in the density field more accurately.

\begin{figure*}
  \begin{center}
   \includegraphics[width=174mm]{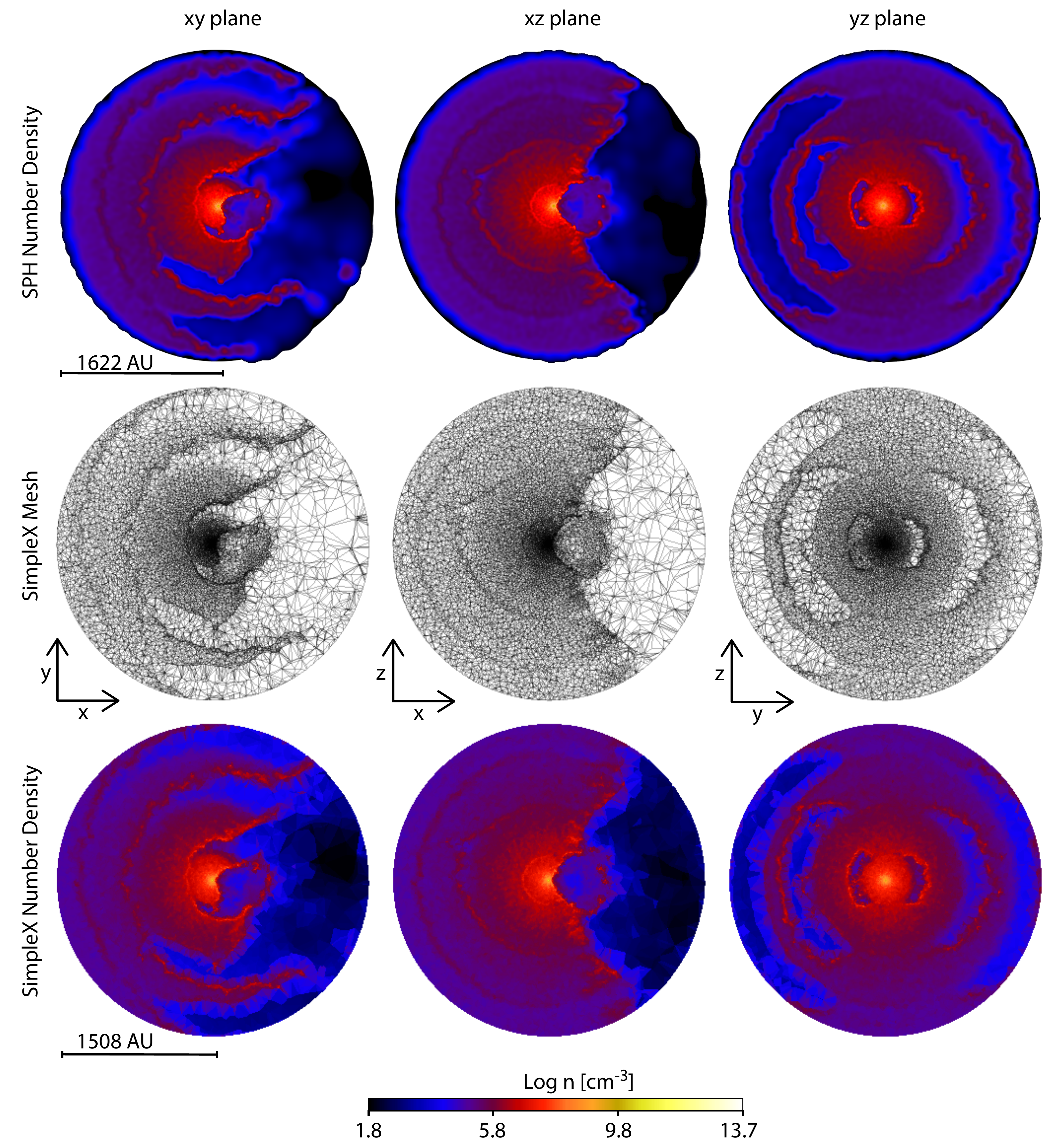}
    \caption{Slices in the $xy$- (left column), $xz$- (middle column) and $yz$- (right column) planes through the 3D simulation volume for the Case~A simulation at apastron. Rows show, from top to bottom, the original SPH number density distribution (log scale, cgs units), the \SimpleX mesh, and the resulting \SimpleX number density (same log scale, cgs units). The resolution of the \SimpleX mesh, as well the number density, follow well the resolution of the original SPH data. In the first column (i.e. the orbital plane) \etaA is to the left and \etaB is to the right. The length scales are shown under the top and bottom left panels. Note that the domain size in the \SimpleX snapshots is slightly smaller than that of the SPH simulations only because we have, for visualization purposes, removed the border points used to generate the \SimpleX mesh.}\label{fig:EtaCarWIND895}
  \end{center}
\end{figure*}

Figure~\ref{fig:EtaCarWIND895} presents an example of the resulting \SimpleX mesh and number density at apastron for a typical 3D SPH simulation of \ec. The first row shows the original number density from the SPH simulation for slices in the $xy$-, $xz$-, and $yz$-planes for the Case~A simulation. The \SimpleX mesh (second row) reproduces everywhere the features present in the original SPH data. The resulting \SimpleX number density (third row) follows extremely well the SPH one in shape, resolution, and value.

\citet{Paardekooper_etA_2010} and \citet{Kruip_etA_2010} discussed how undersampling can have a negative effect on the outcome of an RT simulation that uses grid-based data. In principle, the same problem holds for particle-based data in areas where sharp gradients are present in the number density of SPH particles. To ensure our results are not prone to such issues, we developed a method to increase the resolution of any sparsely sampled regions. We begin with the triangulation of the SPH particles. For every tetrahedron in this triangulation that is larger than a tolerance volume, an additional vertex is placed in the centre of the tetrahedron, and 1/5 of the mass of the four vertices that constitute the tetrahedron is given to the new vertex. This procedure is manifestly mass conserving and regularizes the mesh in low resolution regions.

To study the influence this procedure has on the RT results, we performed test RT simulations with and without resolution enhancements in normally sparsely-sampled regions. For these tests we used a characteristic \ec SPH simulation snapshot in which the densities span roughly ten orders of magnitude (i.e. the top row of Figure~\ref{fig:EtaCarWIND895}). If the RT results are sensitive to sharp gradients, we expect significant differences between the RT simulation with increased resolution and the original non-enhanced simulation.

Although the difference is not large, the overall density in the affected regions is slightly enhanced by the new interpolation. However, we find that the overall shape and spatial extent of the ionized regions does not change. The ionization fraction in the local region most affected by the increased resolution is somewhat lower after the procedure though. This is consistent with the notion that the interpolation increases slightly the density in the higher-resolution region, resulting in a slightly higher recombination rate. The difference, however, is negligible. These results suggest that locally increasing the resolution does not change the overall shape and size of the ionized regions. We are therefore confident that our RT results are not susceptible to systematic effects related to strong gradients in the SPH particle number density \citep[for further detail see Chapters 3 and 4 of][]{Kruip_2011}.


\subsubsection{Ionization State and Chemistry of the Gas}\label{ssec:Chemistry}

In a gas with cross section for photoionization $\sigma( {\vec x}, \nu)$ at position ${\vec x}$, the local photoionization rate, $\Gamma_{P, \ i}({\vec x})$ (which gives the number of photoionizations per second per atom of species $i$ in units [s$^{-1}$]), is given by \citep{Osterbrock_etA_2006}:

\begin{equation}
\label{eq:gamma}
 \Gamma_{P,\ i}({\vec x}) \equiv \int_{0}^{\infty} \frac{4 \pi J_{\nu}({\vec x})}{h \nu} \sigma_{i}({\vec x},\nu) \ {\rm d} \nu,
\end{equation}\\
where $J_{\nu}({\vec x})$ is the local mean intensity and the three species capable of absorbing ionizing photons in the code are \HI, \HeI and \HeII. For simplicity, we only consider the ionization of hydrogen and helium atoms. Implementing ionization processes in a numerical code requires that the relevant equations be expressed in a discretised form. In particular, we need to know the ionization rate per species in each cell of our computational grid. In \SimpleX, ionizing radiation travels from cell to cell along the Delaunay edges. At the nucleus of each Voronoi cell, photons are taken away from the incoming radiation field and their energy is used to ionize the neutral atoms of that Voronoi cell. Given the number densities of these species ($n_{\HI}$, $n_{\HeI}$ and $n_{\HeII}$), and the path length through the cell $l$, the monochromatic optical depth of ionizing radiation $\tau_{\nu}$ is

\begin{equation}
\tau_{\nu}\equiv (n_{\HI} \sigma_{\HI}+ n_{\HeI} \sigma_{\HeI}+ n_{\HeII} \sigma_{\HeII})l.
\end{equation}\\
The total number of ionizations per unit time, $\dot N_{{\rm ion}}$ for a cell with optical depth $\tau_{\nu}$ is then given by

\begin{equation}\label{eq:nion}
\dot N_{{\rm ion}}=\int_{0}^{\infty} \dot N_{\gamma}(\nu)[1-\exp(-\tau_{\nu})]\ {\rm d}\nu,
\end{equation}\\
where $\dot N_{\gamma}(\nu)$ is the number of ionizing photons per unit time streaming into the cell. To quantify how much of the resulting ionizations is due to a particular species, we use the contribution to the total optical depth of that species. The number of ionizations of species $i$ per unit time is

\begin{equation}\label{eq:nioni}
\dot N_{{\rm ion},\ i} =\dot N_{{\rm ion}} \int_{0}^{\infty} \frac{ \tau_{\nu,\ i} }{\tau_{\nu}}\ {\rm d}\nu
\end{equation}\\
Dividing by the number of neutral atoms of species $i$ in the cell, $N_{i}$, then gives the spatially discretised equivalent of Equation (\ref{eq:gamma})

\begin{equation}\label{eq:gamma2}
\Gamma_{P,\ i} = \frac{\dot N_{{\rm ion},\ i}}{N_{i}}.
\end{equation}

In numerical simulations involving radiation it is often necessary to approximate the continuous spectrum of radiation with a finite number of discrete frequency bins due to memory requirements. The extreme (but often employed) limit of one single frequency bin is commonly referred to as the `grey approximation'.
Although in the grey approximation all spectral information is lost, it is still possible to enforce the
conservation of a quantity of importance such as the number of ionizations per unit time or the energy deposition into the medium per unit time. For simplicity, we employ the grey approximation in this work.

The conservation of ionizations is accomplished by defining the effective cross section for species $i$, $\sigma_{I,\ i}$, as

\begin{equation}
\sigma_{I,\ i} = 4 \pi\int^{\infty}_{0}  \frac{\sigma_{i}(\nu) J_{\nu}}{h\nu}\ {\rm d}\nu \big/ \dot N
\end{equation}\\
where $\dot N$ is the rate of ionizing photons per surface area defined by

\begin{equation}\label{eq:ionrate}
\dot N \equiv 4 \pi \int_{0}^{\infty}  \frac{ J_{\nu} }{h\nu}\  {\rm d}\nu.
\end{equation}\\
The photoionization rate is thus given by

\begin{equation}\label{eq:ionrate2}
\Gamma_{P,\ i} = \sigma_{I,\ i} \dot N.
\end{equation}

In addition to photoionization, we include collisional ionization due to the interaction of free electrons and neutral atoms. As this is a kinetic process, the collisional ionization rate, $\Gamma_{C}$, depends on the thermal state of the electrons and is given by

\begin{equation}
\Gamma_{C} =n_{e} \sum_{i} \Gamma_{i}(T) n_{i},
\end{equation}\\
where  $\Gamma_{i}(T)$ are the collisional ionization rates and $n_{e}$ is the electron number density. The total ionization rate is the sum of photo- and collisional ionization rates, $\Gamma = \Gamma_{P}+\Gamma_{C}$.

The inverse process of ionization is recombination. This free-bound interaction of electrons and ions depends on temperature and number density of ions and electrons. The number of recombinations per unit time per hydrogen atom $[\mathrm{s}^{-1}]$ is
\begin{equation}\label{eq:recrate}
  R_{i} = n_{{\rm e}} \alpha_{i}(T),
\end{equation}\\
where $\alpha_{i}(T)$ is the recombination coefficient of species $i$.

We note that we use the `case~B' recombination coefficient where the recombination transition to the ground-state is excluded under the assumption that the radiation associated with this transition is absorbed nearby, resulting in a new ionization. This is referred to as the `on-the-spot' approximation. Details on the implementation of these processes, as well as the various rates and cross sections used can be found in Chapter~5 of \citet{Kruip_2011}.

Together, ionizations and recombinations determine the ionization-state of the gas, described by the following three coupled differential equations and three closure relations

\begin{eqnarray}\label{eq:rate}
\dot n_{ \HI } & = &  n_{\HII}R_{\HII} - n_{\HI} \Gamma_{\HI} \nonumber \\
\dot n_{ \HeI} & = & n_{\HeII}R_{\HeII} - n_{\HeI} \Gamma_{\HeI} \nonumber \\
\dot n_{ \HeII } & = & n_{\HeIII}R_{\HeIII} - n_{\HeII} \Gamma_{\HeII}\\
n_{{\rm H}}&=& n_{ \HI}+ n_{\HII } \nonumber\\
n_{{\rm He}}&=& n_{ \HeI}+ n_{ \HeII}+n_{\HeIII} \nonumber \\
n_{e} & = & n_{\HII} + n_{\HeII} + 2 n_{\HeIII} \nonumber .
\end{eqnarray}\\
This set of equation does not have a general analytical solution and must be solved numerically. For this purpose we adopt the sub-cycling scheme described in \citet{Pawlik_etA_2008}. In this scheme, ionizations and recombinations are evolved on a timescale that is smaller than the ionization or recombination timescales $t_{{\rm ion}}$ and $t_{{\rm rec}}$. During a radiative transfer time-step, the ionizing flux is assumed to be constant, making the procedure manifestly photon-conserving. This allows for radiative time-steps $\Delta t_{{\rm rt}}$ that are much larger than the dominant timescale governing the evolution of the ionization state. The sub-cycling time-step for both ionization and recombination is

\begin{equation}\label{eq:tsub}
\Delta t_{{\rm sub}} \equiv \frac{t_{{\rm ion}}t_{{\rm rec}}}{t_{{\rm ion}}+t_{{\rm rec}}}.
\end{equation}\\
Because the procedure is analogous for each species, we give here only the explicit example for the integration step for hydrogen. At time $t_{{\rm sub}} \in \left( t_{{\rm rt}}, t_{{\rm rt}} + \Delta t_{{\rm rt}}  \right)$ the rate equation is given by

\begin{equation}\label{eq:subcycle}
    \mbox{d}n_{ {\rm H\,II} }^{(t_{{\rm sub}})} = n_{{\rm  H\,I}}^{(t_{{\rm sub}})} \Gamma_{{\rm  H}}^{(t_{{\rm sub}})} \Delta t_{{\rm sub}} - n_{{\rm e}}^{(t_{{\rm sub}})} n_{{\rm  H\,II}}^{(t_{{\rm sub}})} \alpha_{{\rm  H}}(T) \Delta t_{{\rm sub}},
\end{equation}\\
where the photoionization rate at $t_{{\rm sub}}$ is

\begin{equation}\label{eq:subcycle2}
    \Gamma_{{\rm  H}}^{(t_{{\rm sub}})} = \Gamma_{{\rm  H}} \left( \frac{ 1-e^{- \tau^{(t_{{\rm sub}})}} }{ 1-e^{- \tau} } \right) \frac{ n_{{\rm  H\,I}} }{ n_{{\rm  H\,I}}^{(t_{{\rm sub}})} },
\end{equation}\\
where $\Gamma_{{\rm  H}}$ and $\tau$ are the photoionization rate and optical depth at the beginning of the sub-cycling and $\tau^{(t_{{\rm sub}})} = \tau \, n_{{\rm  H\,I}}^{(t_{{\rm sub}})} / n_{{\rm  H\,I}} $. By defining the photoionization rate in this way, the ionizing flux in the cell is constant during the radiative transfer time-step. This sub-cycling scheme becomes computationally expensive when $\Delta t_{{\rm sub}} \ll \Delta t_{{\rm rt}}$, but photoionization equilibrium is generally reached after a few sub-cycles. It is then no longer necessary to explicitly integrate the rate equation, but instead use the values of the preceding sub-cycle step. This way of sub-cycling ensures photon conservation even for large radiative transfer time-steps.


\subsection{Application of \SimpleX to \ec}\label{ssec:EtaCar}

Since the \SimpleX calculations are performed as post-processing on the 3D SPH simulation output, we use snapshots corresponding to an orbital phase of apastron (Figure~\ref{fig:EtaCarWIND895}). The reason for this choice lies in the slow dynamical changes that the system undergoes around apastron. This `stable' situation allows us to run RT simulations for a sufficiently long time without worrying about important changes to the 3D structure of the system that occur around periastron (\citealt{Okazaki_etA_2008, Parkin_etA_2011}; \citetalias{Madura_etA_2012, Madura_etA_2013}). Moreover, the \emph{HST}/STIS mapping data currently in-hand to be modeled was taken at phases around apastron during \ec's orbital cycle \citep{Gull_etA_2011, Teodoro_etA_2013}. Detailed modeling of future (late 2014 through early 2015) \emph{HST} observations obtained across \ec's periastron event is deferred to future work.

We focus on the ionization of H and He due to \etaB, assuming the same abundance by number of He relative to H as \citet{Hillier_etA_2001}, $n_{\rm He}/n_{\rm H}=0.2$. The reasons for this single-source approximation are discussed in Section~\ref{ssec:Primary}. We performed tests to determine the correct time-step for accurate RT calculations of the ionization volumes and fractions, and find that a simulation time-step of $\sim 3$~s is required. The SPH output is post-processed with \SimpleX until the ionization state reaches an equilibrium value. This typically happens within  $\sim 3$ months for the SPH snapshots investigated. We thus set the total \SimpleX simulation time to 3~months. Because the gas is assumed to be initially almost fully neutral, this provides an upper limit on the time it takes before convergence is reached. Since this limit is well within the orbital time-scale around apastron, this is another indication that post-processing of the SPH simulations does not significantly alter our results.


\subsubsection{Influence of the Primary Star \etaA}\label{ssec:Primary}


Detailed fitting of the optical and UV spectra of \ec by \citet{Hillier_etA_2001,Hillier_etA_2006} and \citet{Groh_etA_2012} shows that for $\MdotA \approx 8.5 \times 10^{-4} \Msy$, the region of fully ionized H around \etaA extends radially $\sim120$~AU, while the region of doubly-ionized He extends $\sim 0.7$~AU, and that of singly-ionized He from $\sim 0.7$ to $3$~AU. Assuming a constant spherical mass-loss rate, the density in the \etaA wind is expected to fall off as $r^{-2}$. To explore the dependence of the position of the ionization fronts on the ionizing luminosity of \etaA, we performed 1D calculations using an equilibrium chemistry solution where the ionization fractions of hydrogen and helium are set to their equilibrium values under the assumption that the incoming flux of ionizing photons is constant. As mentioned in Section~\ref{ssec:Chemistry}, the ionization state of the gas is described by the first three equations in set (\ref{eq:rate}). We can derive the equilibrium equations by setting $\dot n_{ \HI} =\dot n_{ \HeI} =\dot n_{ \HeII}=0$. After some algebra this yields

\begin{eqnarray}\label{eq:equilibrium}
x_{\HI} &= & \left( 1 + \Gamma_{\HI}/R_{\HII} \right)^{-1} \nonumber \\
x_{\HII} & = & 1 - x_{\HI} \nonumber \\
x_{\HeI} &= & \left[ 1 + \Gamma_{\HeI}/R_{\HeII} \times \left( 1+ \Gamma_{\HeII}/R_{\HeIII} \right) \right]^{-1}\\
x_{\HeII} &= & x_{\HeI}  \Gamma_{\HeI}/R_{\HeII} \nonumber \\
x_{\HeIII} &= & x_{\HeII}  \Gamma_{\HeII}/R_{\HeIII} \nonumber ,
\end{eqnarray}\\
where $x_{i}$ is the fraction of species $i$ and we have used $n_{i} = x_{i} n_{{\rm j}}$ where $j \in ({\rm H}, {\rm He})$. These equations are coupled by the free electron density given by the last equation in (\ref{eq:rate}).

Unfortunately, the set of equations (\ref{eq:equilibrium}) cannot be solved analytically due to the non-linear dependence on ionization fractions of the photoionization rate through the optical depth. More specifically, the photoionization rate in a cell is given by Equation~(\ref{eq:gamma2}), where the monochromatic analog of Equation~(\ref{eq:nion}) is

\begin{equation}
\dot N_{\mathrm{ion}}= \dot N_{\gamma}[1-\exp(-\tau)],
\end{equation}\\
with $\tau=(x_{\HI} n_{ {\rm H} } \sigma_{ \HI}  + x_{\HeI} n_{ {\rm He} } \sigma_{ \HeI } +x_{\HeII} n_{ {\rm He} } \sigma_{ \HeII} ) l$. Because of this non-linear dependence, the equilibrium fractions must be solved iteratively. If the iterative procedure converges, the neutral fractions are assigned to the cell under treatment and the flux is diminished by the number of absorptions during that time-step $\Delta t$ ($n_{\HI} \Gamma_{\HI} \Delta t $).

The density profile used in the 1D code is obtained from the expression $\rho (r) =\MdotA/4 \pi r^2 v(r)$, where  $v(r) = v_{\infty}(1-R_{\star}/r)^{\beta}$ (see M13, eqn.~A1). The 1D code simulates radiation traveling through spherically symmetric shells with a maximal radius of the simulation, $\approx 1622$~AU. The radiation is injected in the first shell and then travels outward until it is either absorbed or exits the last shell. For the results shown we used $\sim 3 \times 10^{4}$ shells. We note that time-stepping is arbitrary because of the equilibrium chemistry. The only variable is therefore the luminosity of the source \etaA.

\begin{figure}
  \begin{center}
   \includegraphics[width=84mm]{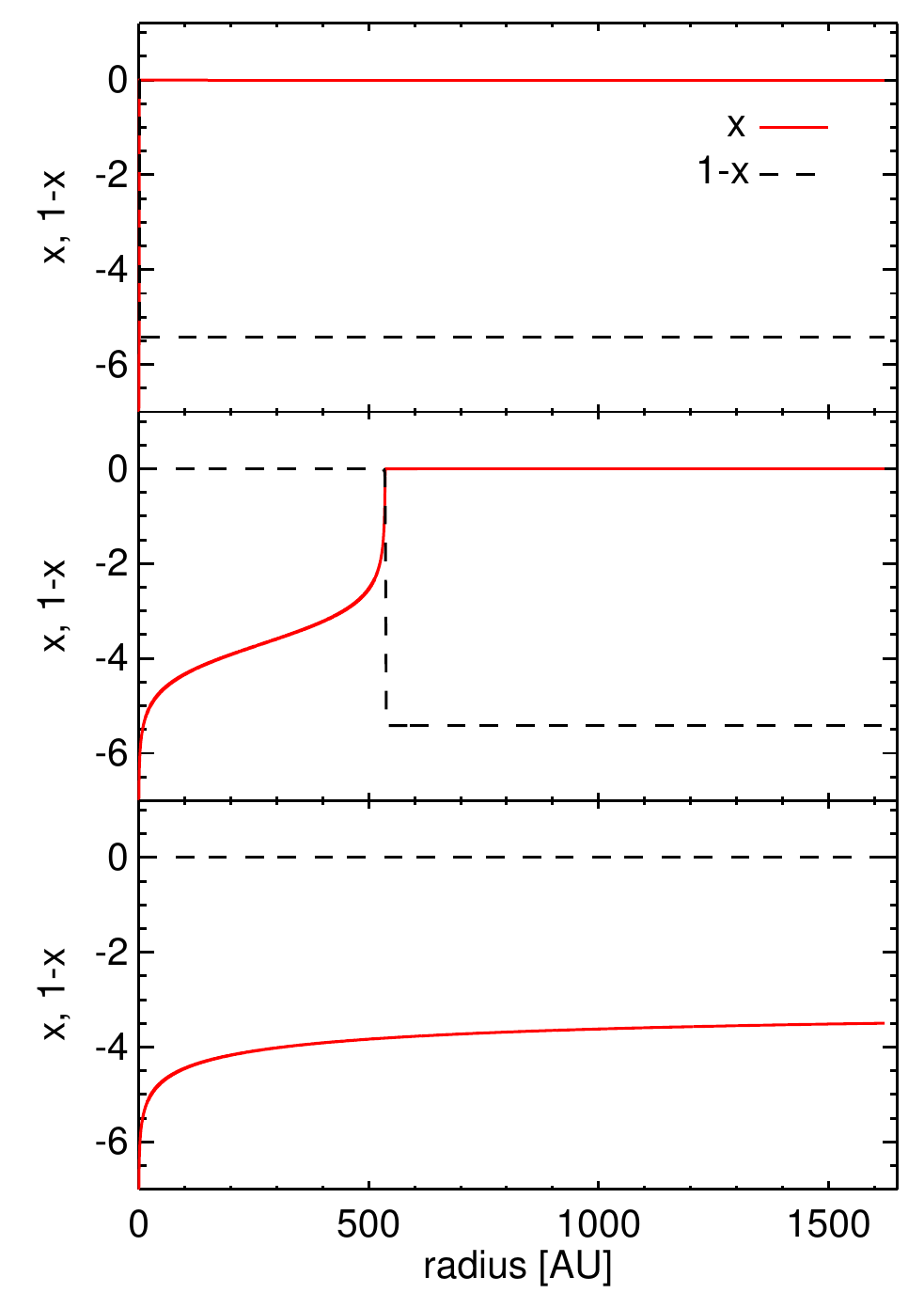}
   \caption{Neutral (\textsf{x}, solid lines) and ionized ($1-$~\textsf{x}, dashed lines) fractions of H (log scale) as a function of distance for the 1D RT simulations of \etaA described in Section~\ref{ssec:Primary} and luminosities of \emph{Top}: $1.9 \times 10^{53}~{\rm s^{-1}}$, \emph{Centre}: $ 1.91756 \times 10^{53}~{\rm s^{-1}}$, and \emph{Bottom}: $1.91758 \times 10^{53}~{\rm s^{-1}}$. This behaviour signifies that the H ionization front is highly unstable and thus changes qualitatively with small perturbations in either the density or the luminosity.}\label{fig:primary}
  \end{center}
\end{figure}

Using $\MdotA \approx 8.5 \times 10^{-4} \Msy$, for luminosities below a critical value ($1.9 \times 10^{53}~{\rm s^{-1}}$), the computational box is neutral and the Str\"omgren radii are confined to the central 0.7~AU (top panel, Figure~\ref{fig:primary}). The H ionization front is located somewhere between the centre and the outside of the box for a very small range of luminosity values (centred around $1.91756 \times 10^{53}~{\rm s^{-1}}$, centre panel, Figure~\ref{fig:primary}). The slightest increase in luminosity results in a completely ionized box, while further increases result only in a lower neutral fraction throughout the simulation volume (bottom panel, Figure~\ref{fig:primary}). This behaviour is completely expected, however, for ionization fronts in power-law density profiles with powers less than $-2/3$ \citep{Franco_etA_1990, Shapiro_etA_2006}. For such profiles, the circumstellar medium simply cannot support stable ionization fronts.

For these reasons, constraining the ionization fronts in \etaA 's wind to the values derived by \citet{Hillier_etA_2001,Hillier_etA_2006} and \citet{Groh_etA_2012} using \SimpleX is practically impossible given the fronts' intrinsically unstable nature. We realize this 1D result is over-simplified since the instability is real in a pure H or H $+$ He gas, but will disappear with the introduction of the myriad of spectral lines, mostly by Fe, that have a so-called `line blanketing' effect on the stellar spectra. However, the inclusion of such additional species and blanketing effects is beyond the scope of this paper.

Given the above difficulties, the most sensible choice for an initial effort to model the ionized WWIRs is to omit \etaA's radiation altogether. This may seem an oversimplification at first, but there are several arguments for this approximation. First, for the high \MdotA Case~A and B simulations, 1D \texttt{CMFGEN} models by \citet{Hillier_etA_2001,Hillier_etA_2006} show that the primary source will sustain an ionized hydrogen region that spans roughly 240--260~AU in diameter, $\sim 0.04\%$--0.05\% of the volume of the SPH and \SimpleX simulations. The same models show that the total sum volume of singly- plus doubly-ionized helium in the central \etaA wind accounts for $\sim 10^{-6}$\% and $10^{-5}$\% of the \SimpleX simulation volume for Cases~A and B, respectively. These volumes are too close to the primary and too small to directly affect the ionization fronts and fractions at the locations where the spatially-extended, high-velocity forbidden line emission forms, especially at orbital phases near apastron (\citealt{Gull_etA_2009, Gull_etA_2011}; \citetalias{Madura_etA_2012}; \citetalias{Madura_etA_2013}). They may, however, influence the ionization structure further away indirectly by reducing the opacity for photons from the secondary source. This may be especially true very close to periastron. We expect that this would primarily result in UV flux from \etaA penetrating the WWIR more easily, effectively increasing the ionized fraction on the far side of the primary source. For an observer on Earth though, this region is, at periastron, located behind \etaA and therefore likely obscured by the dense primary wind.

Second, although extremely luminous, because it is enshrouded by a dense, optically-thick wind, \etaA has a spectrum representative of a much cooler star than \etaB \citep{Hillier_etA_2001, Hillier_etA_2006}. The effective temperature of \etaA at optical depth $\tau = 2/3$ ($r \approx 4$~AU) is predicted to be $\sim 9200$~K for Cases~A and B \citep{Hillier_etA_2001, Hillier_etA_2006}. The ionizing flux from \etaA is thus substantially diminished before reaching the WWIR, located $\sim$ 20--22~AU from \etaA when the system is near apastron. Since essentially zero photons with energies above 13.6~eV from \etaA reach the WWIRs on the apastron side of the system at times near apastron, omission of the \etaA source is a justifiable simplification when the focus is on forbidden emission lines with ionization potentials above 13.6~eV.

One might try to argue that because the ionized hydrogen and helium volumes in the inner primary wind extend far enough to encompass both stars and the WWIR at phases close to periastron, photons from \etaA will also reach the apastron side of the simulation volume. This argument relies, however, on the assumption that the ionized regions are indeed spherical and therefore penetrate the WWIR toward the secondary star. This assumption is likely incorrect given the high density of the WWIR. In other words, we would be applying a model based on spherical symmetry to a region that clearly has a very asymmetrical geometry.

The exception to the above arguments is the Case~C simulation. In this instance, according to 1D \texttt{CMFGEN} models (\citealt{Hillier_etA_2006}; \citetalias{Madura_etA_2013}), H is fully ionized in the pre-shock \etaA wind throughout the entire simulation domain. Moreover, the \HeII region in the inner \etaA wind extends radially $\sim 120$~AU. While neglecting the \etaA ionizing source in this case likely produces incorrect RT results in the regions of \etaA wind on the periastron side of the system, for our purposes, the situation is actually not so bad. First, we note that, based on previous works, the mass-loss rate of \etaA is very likely not as low as the value assumed in the Case~C situation (see arguments in e.g. \citealt{Hillier_etA_2006, Parkin_etA_2009, Teodoro_etA_2012, Teodoro_etA_2013, Russell_2013}; \citetalias{Madura_etA_2012}; \citetalias{Madura_etA_2013}), and so we will not be using the \SimpleX results obtained here for Case~C to model the observed broad, high-ionization forbidden line emission. Rather, in addition to investigating how a reduced \MdotA affects the ionization structure on the apastron side of the system, we use Case~C as an illustrative example to determine whether \etaB's ionizing radiation can penetrate the dense WWIRs and further affect the ionization state of \etaA's wind. Having H initially ionized in the pre-shock \etaA wind would primarily influence the ionization structure indirectly by reducing the opacity for photons from \etaB (assuming they can penetrate the WWIR), increasing the ionized fraction of H in the pre-shock \etaA wind, but having little effect on the overall ionization volume.

Regarding the He ionization structure in Case~C, because the inner \HeII region extends $\sim 120$~AU, the innermost WWIR penetrates \etaA's \HeII zone, even at apastron. However, the total volume of this inner \HeII region is still only $\sim 0.04\%$ of the total \SimpleX simulation volume, again too small to directly affect the ionization fronts and fractions at the locations where the high-ionization forbidden lines of interest form. Assuming He is neutral in the \etaA wind at the start of the \SimpleX simulations also allows us to more easily determine whether He-ionizing photons from \etaB can penetrate the WWIRs and affect the pre-shock \etaA wind. If so, the primary effect will be an increased fraction of \HeII in the innermost \etaA wind, with little to no effect on the shape or extent of the \HeII ionization volume. Moreover, since \etaB is thought to be an O- or WR-type star with $T_{\mathrm{eff}} \simeq$ 36,000--41,000~K \citep{Verner_etA_2005, Hillier_etA_2006, Teodoro_etA_2008, Mehner_etA_2010}, the number of photons it produces capable of ionizing \HeII to \HeIII is effectively zero. Thus, there will be no \HeIII region created by \etaB beyond the WWIR zone, even if \etaB's radiation can penetrate the dense post-shock gas. Therefore, even for simulation Case~C, the neglecting of \etaA's radiation is a justifiable simplification for the purposes of our work. The only caveat is that we do not account for any possible ionization of \etaA's pre-shock wind to \HeIII by soft X-rays produced in the $420$ \kms \etaA shock. However, any such \HeIII region near the WWIR zone at times around apastron is very likely to be negligible in extent, if it exists at all, as evidenced by the absence of any significant detectable \HeII $\lambda 4686$ emission in \ec during its spectroscopic high state \citep{Mehner_etA_2011, Teodoro_etA_2012}.


Based on the above considerations, we neglect the \etaA ionizing source in this work and focus on the influence of \etaB.


\begin{figure}
  \begin{center}
   \includegraphics[width=84mm]{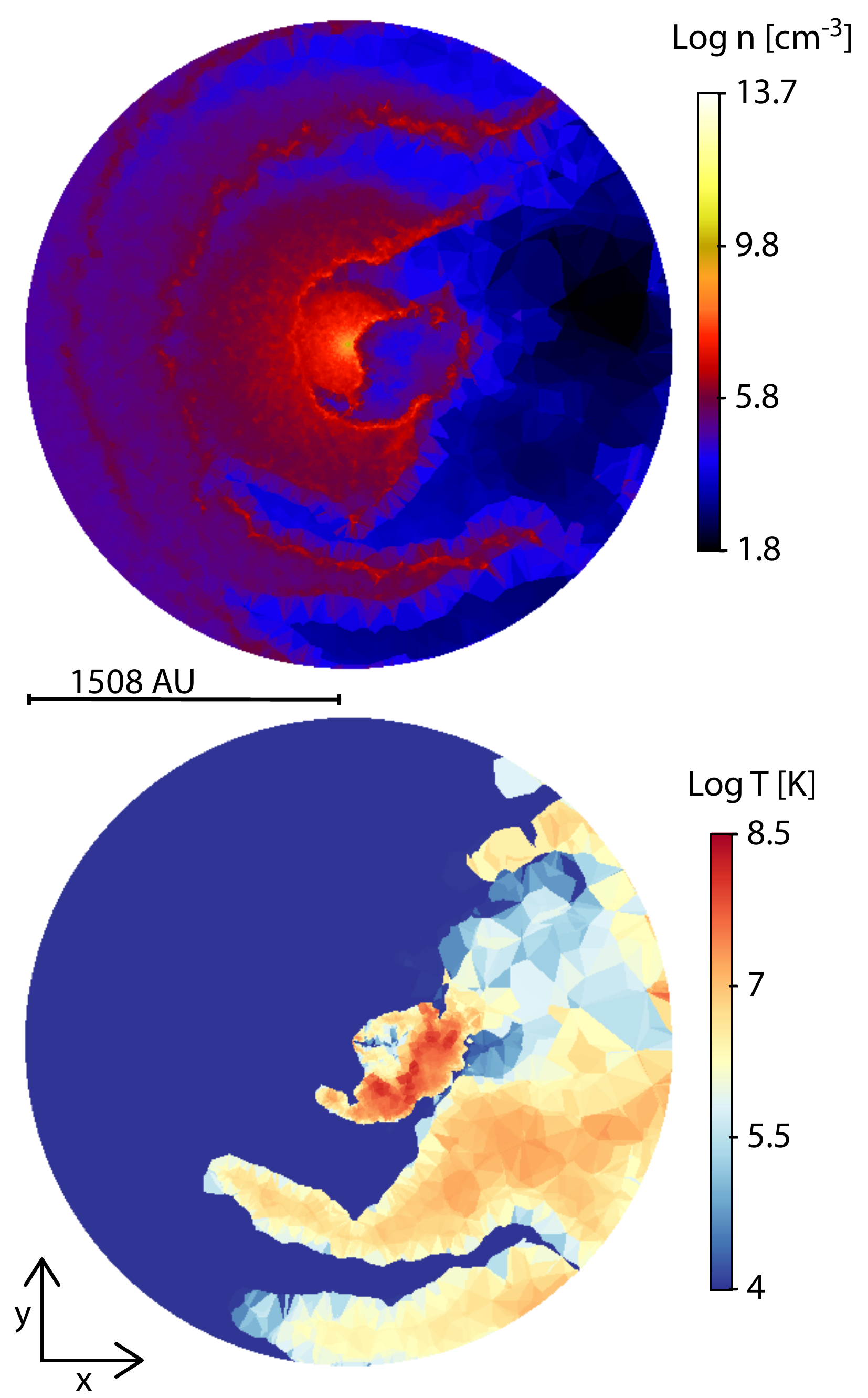}
    \caption{Slices in the $xy$ orbital plane through the 3D simulation volume for the Case~A simulation at apastron. The top plot shows the number density (log scale, cgs units), while the bottom shows log temperature (K). The lower-density \etaB wind is shock-heated by the wind-wind interaction to temperatures up to $\sim 10^{8.5}$~K. On the periastron (left) side the denser \etaA wind and compressed shells formed at periastron radiatively cool to $T \sim 10^{4}$~K.}\label{fig:dens_temp_895}
  \end{center}
\end{figure}

\subsubsection{The Ionizing Source \etaB}\label{ssec:Source}

For the RT calculations, we place a spherical ionizing source centred at the location of \etaB. This `source' is composed of a series of individual points randomly distributed about the sphere that defines the injection radius used in the SPH simulations for the wind of \etaB ($30$~\Rs). We use a total of 50 source points, which is large enough to result in a nearly isotropic photoionizing source. We find that using more points has little effect on the RT results. The total luminosity is divided among all 50 points, forming the nodes of the grid that emit radiation. These nodes are also capable of absorbing any radiation emitted by neighbouring points in the \SimpleX grid. Based on the work of \citet{Mehner_etA_2010, Verner_etA_2005} and M12, we assume for \etaB a total ionizing flux for hydrogen and helium of $3.58 \times 10^{49}$ photons$\,{\rm s}^{-1}$ ($3.02 \times 10^{49}$ capable of ionizing \HI and $5.62 \times 10^{48}$ for ionizing \HeI), consistent with an O5 giant with ${\rm T_{eff}}\approx 40,000$~K \citep{Martins_etA_2005}.

\section{Results}\label{sec:Results}

To provide context for interpreting the RT results below, we begin with a brief description of the density and temperature structure of the system in the orbital plane for the Case~A simulation (Figure~\ref{fig:dens_temp_895}). Example number density slices in the $xz$ and $yz$ planes, plus number density slices in each plane for the Case~B and C simulations, can be found in Figures~\ref{fig:dens_ionH_xy}, \ref{fig:dens_ionH_xz}, and \ref{fig:dens_ionH_yz}. Example slices showing temperature for each case can be found in \citetalias{Madura_etA_2013}. We focus on the Case~A simulation as the \MdotA assumed in this case most likely represents \etaA's current observed mass-loss rate (\citealt{Groh_etA_2012}; \citetalias{Madura_etA_2013}).

\begin{figure*}
  \begin{center}
   \includegraphics[width=174mm]{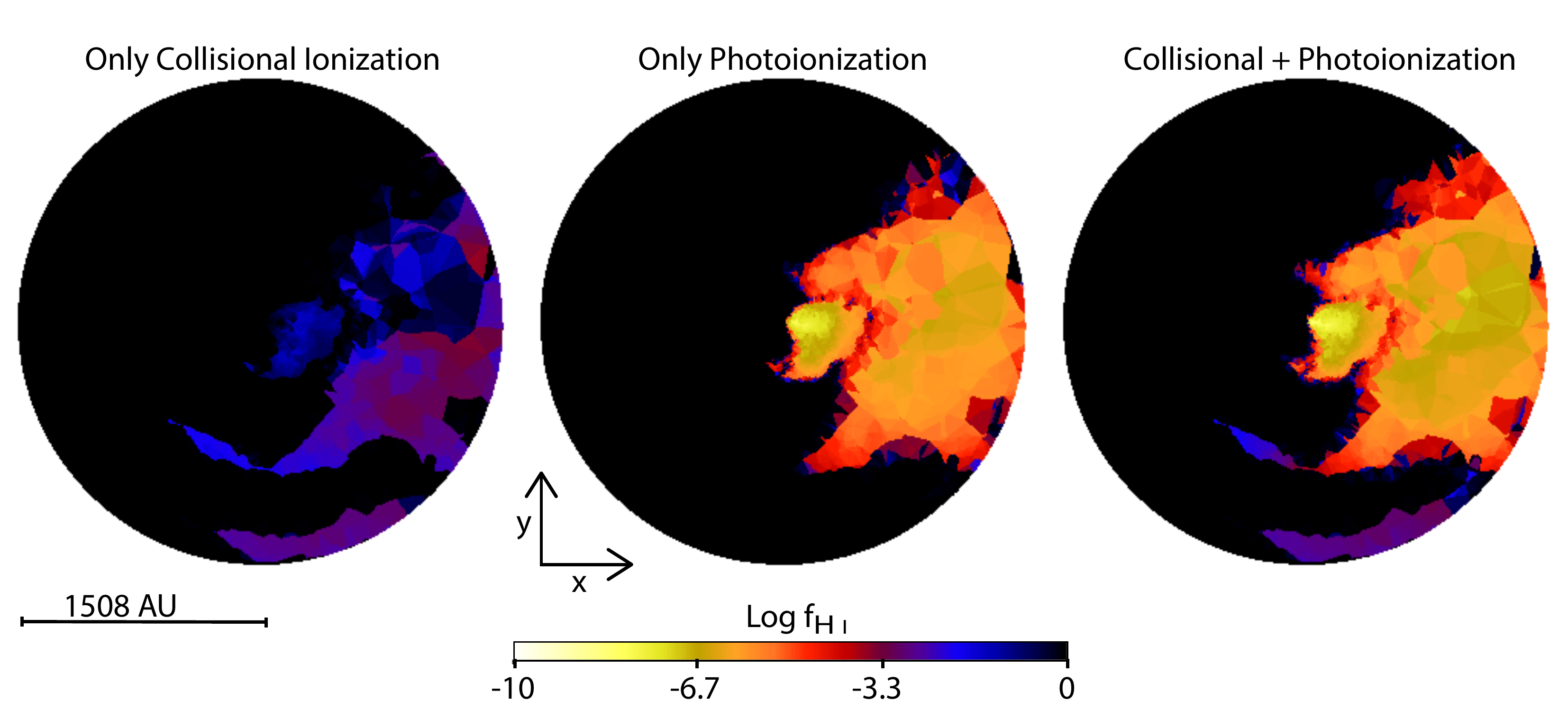}
    \caption{Slices in the $xy$ orbital plane through the 3D \SimpleX simulation volume for Case~A at apastron showing the logarithm of the fraction of neutral hydrogen assuming \emph{Left}: collisional ionization only, \emph{Centre}: photoionization only, and \emph{Right}: collisional- and photo-ionization. Including collisional ionization is necessary to ionize the small cavities in the primary wind and the `fingers' of low-density \etaB wind trapped between the higher-density walls of \etaA wind that form around periastron.}\label{fig:photo_coll_ion_895}
  \end{center}
\end{figure*}

\citet{Pittard_etA_1998, Pittard_etA_2002, Okazaki_etA_2008, Parkin_etA_2009, Parkin_etA_2011, Madura_etA_2012}; and \citet{Russell_2013} showed that \etaB is between the observer and \etaA at apastron. Due to the highly eccentric binary orbit, \etaB spends most of its time near apastron (right side of panels in Figure~\ref{fig:dens_temp_895}), so that the relatively undisturbed wind of \etaA is located on the far (periastron) side of the system. Across every periastron passage, the hot and low density wind of \etaB pushes outward into the slow, high-density \etaA wind, leading to the formation of a thin, high-density wall surrounding the lower-density, trapped wind of \etaB (\citealt{Parkin_etA_2011}; \citetalias{Madura_etA_2013}). This dense wall is accelerated to a velocity higher than the normal terminal velocity of \etaA's wind and expands creating a thin, high-density sheet of trapped primary wind material. During periastron passage the arms of the WWIR  become extremely distorted by orbital motion as the binary stars move toward their apastron positions. Moving back toward apastron, orbital speeds decrease and the \etaB wind cavity regains its axisymmetric conical shape (\citealt{Okazaki_etA_2008, Parkin_etA_2011}; \citetalias{Madura_etA_2013}).

Dense arcs and shells of \etaA wind visible in the outer regions on the apastron side of the system in the top panel of Figure~\ref{fig:dens_temp_895} highlight the fact that the binary has already undergone multiple orbits. Narrow cavities carved by \etaB\ in \etaA's dense wind during each periastron passage also exist on the periastron side of the system. Bordering these narrow cavities are the compressed, density-enhanced shells of primary wind formed as a result of the rapid wind-wind collision during each periastron.

While the periastron side of the system is dominated by the dense wind of \etaA, the apastron side is dominated by the much lower-density, faster wind of \etaB, although arcs of compressed \etaA wind also extend to the apastron side. These arcs are the remnants of the shells of \etaA wind that flow in the apastron direction when \etaB is at periastron \citepalias{Madura_etA_2013}. The partially intact, most recent shell is visible just to the right of the centre of the image in the top panel.

There is also a clear temperature asymmetry between the apastron and periastron sides of the system (bottom panel of Figure~\ref{fig:dens_temp_895}). The gas on the periastron side is much colder at $T \approx 10^{4}$~K. The various wind-wind collisions on the apastron side produce large volumes of gas shock-heated to temperatures between $10^{6}$ and $10^{8.5}$~K. Because the gas on the apastron side is composed mostly of \etaB wind material of low density, it cools slowly and adiabatically, allowing it to remain hot throughout the 5.54-year orbital cycle. In contrast, the dense shells of post-shock primary wind cool radiatively very quickly down to $T \sim 10^{4}$~K (\citealt{Parkin_etA_2011}; \citetalias{Madura_etA_2013}). The innermost region of the system where the current WWIR is located, and the region where \etaB's wind collides with the latest ejected shell of primary wind, have the highest temperatures and are responsible for the observed time-variable X-ray emission \citep{Hamaguchi_etA_2007, Okazaki_etA_2008, Corcoran_etA_2010, Parkin_etA_2011}.


\subsection{The Importance of Collisional Ionization}\label{ssec:Coll}

The bottom panel of Figure~\ref{fig:dens_temp_895} shows that the shocks induced by the violent wind-wind collisions heat the gas in the system to very high temperatures. Since the lower-density material from \etaB cools adiabatically, this gas remains extremely hot for most of the orbit, at temperatures well above those where collisional ionizations become important ($\gtrsim 10^6$~K). The collisional ionization fraction depends strongly on the temperature of the gas, which is in principle a function of the hydrodynamical motion, photo-heating, and multiple cooling terms. In this initial study, as a first approximation we use the temperature calculated by the SPH code to estimate the importance of collisional ionizations. In order to assess which process dominates, we performed a series of simulations with/without collisional-/photo-ionization. For brevity, we discuss here only the results for hydrogen for simulation Case~A. Results for helium and Cases~B and C are qualitatively similar.

Figure~\ref{fig:photo_coll_ion_895} summarizes the results. The three panels represent the \SimpleX output if we include, respectively, only collisional ionization, only photoionization, or both. For the only collisional ionization case, we assume collisional ionization equilibrium as an initial condition to the RT, as described in Section~\ref{ssec:Chemistry}. In this case, the overall ionization structure unsurprisingly follows the plot of the temperature in Figure~\ref{fig:dens_temp_895}. The cold, dense primary wind on the periastron side of the system, and the dense WWIRs of compressed primary wind that extend to the apastron side of the system (both in black in the first panel of Figure~\ref{fig:photo_coll_ion_895}), remain mostly neutral. However, hydrogen in the hot, lower-density regions of shocked secondary wind are collisionally ionized (in blue and purple in the first panel of Figure~\ref{fig:photo_coll_ion_895}). The \etaB wind in the outermost parts of the system is the most highly ionized due to the much lower density of the gas there, which results in less recombination. We also note in particular the two `fingers' of highly ionized \etaB gas that extend into the primary wind, located at the bottom of the panel. More importantly, we see that when only collisional ionizations are used, the dense WWIRs remain almost entirely neutral. 

In the case of only photoionizations from \etaB (middle panel of Figure~\ref{fig:photo_coll_ion_895}), mainly the lower-density \etaB wind on the apastron side of the system is highly ionized (in yellow and orange). The lower-density hot fingers of \etaB wind trapped between the high-density walls of \etaA wind show no ionization. These regions are effectively shielded from the ionizing flux of \etaB. Another important difference is the level of ionization in the \etaB wind. Photoionizations are capable of reducing the fraction of neutral hydrogen by roughly four more orders of magnitude, compared to the case with only collisional ionization. Additionally, the \etaB wind closest to the centre of the simulation is the most highly ionized since, even though the density is higher there, the material is much closer to the luminous ionizing source. This is the exact opposite of what was observed in the case of only collisional ionizations. We also see that photons from \etaB are capable of penetrating the innermost wall of \etaA wind material on the apastron side of the system, thus also highly ionizing it and the outer portions of \etaB's wind. Detailed examination further shows that when photoionizations are used, the edges of the WWIRs facing \etaB can be significantly ionized ($\log f_{\HI} \lesssim -3$, Madura \& Clementel 2014, in prep.).

Using both collisional- and photo-ionizations results in a situation that resembles a superposition of the first two panels (right panel of Figure~\ref{fig:photo_coll_ion_895}). The \etaB wind on the apastron side remains highly ionized, but collisional ionization helps ionize the fingers of \etaB wind located at the bottom of the panel. Interestingly, the \HI in the fingers is slightly more ionized now compared to the case with only collisional ionizations. This is because, due to the now reduced opacity caused by including collisional ionization, photons from \etaB can more easily penetrate into the fingers and increase the overall level of ionization. A similar effect is seen in/near the WWIRs, which are also slightly more ionized when both collisional- and photo-ionizations are included, compared to the case with only photoionization. However, even when both collisional- and photo-ionizations are used, the dense \etaA wind on the periastron side of the system remains neutral.


Given all of these results, we consider both collisional- and photo-ionizations as necessary for any proper RT simulations of \ec. The remainder of the results in this paper are based on simulations that accordingly incorporate both.


\subsection{Overall Ionization Structure and Influence of \MdotA}\label{ssec:Ion&Mdot}

\subsubsection{The orbital plane}

\begin{figure*}
  \begin{center}
   \includegraphics[width=174mm]{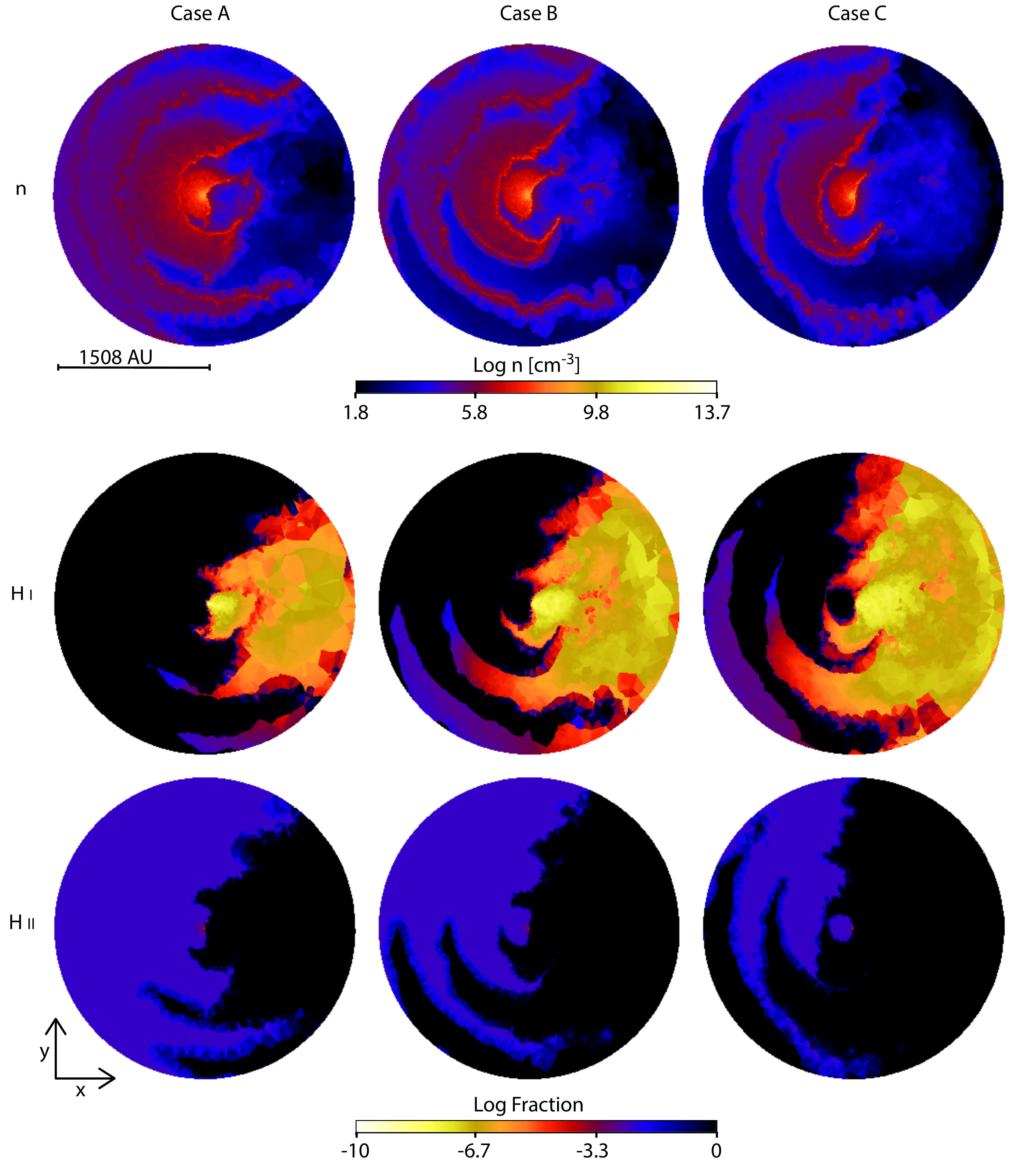}          
\caption{Slices in the $xy$ orbital plane through the 3D \SimpleX simulation volume for the three different assumed \MdotA. Columns illustrate, from left to right, Cases A to C. Rows show, from top to bottom, the \SimpleX number density (log scale, cgs units) and the computed fractions of \HI and \HII (log scale).}\label{fig:dens_ionH_xy}
  \end{center}
\end{figure*}

The top row of Figure~\ref{fig:dens_ionH_xy} shows the \SimpleX number density in the orbital plane for simulation Cases~A--C. As demonstrated by \citetalias{Madura_etA_2013}, \MdotA determines the overall shape of the WWIRs and the stability of the arcuate shells expanding on the apastron side of the system. Lowering \MdotA increases the opening angle of the shock cone created by \etaB, increasing the volume of low-density \etaB wind. This is particularly noticeable in the size of the low-density fingers of \etaB wind that strongly reduce the volume of unperturbed primary wind. The lower the \MdotA, the wider and more extended the fingers. In Cases~B and C, the fingers extend to the back (periastron) side of \etaA's wind. The dense shells of \etaA wind on the apastron side are also more stable and remain intact longer for higher values of \MdotA \citepalias{Madura_etA_2013}. As a consequence, we expect that the 3D shape, position, intensity, and variability of the ionization depend strongly on \MdotA.

The middle and bottom rows of Figure~\ref{fig:dens_ionH_xy} present, respectively, the computed fractions of \HI and \HII in the orbital plane. The WWIRs and high-density walls surrounding the lower-density trapped wind of \etaB define the separation between the neutral and ionized-hydrogen regions. These high-density \etaA wind structures are able to trap the hydrogen ionizing photons from \etaB. We also see that as \MdotA decreases, the volume of ionized hydrogen increases greatly on both the apastron and periastron sides of the system. The larger fingers for Cases~B and C allow the ionizing radiation from \etaB to penetrate into the low-density cavities that are carved within the back side of the primary wind every periastron passage.

In the \HII maps of Figure~\ref{fig:dens_ionH_xy} it is possible to see a large fraction of neutral hydrogen at and to the periastron side of \etaA. As described in Section~\ref{ssec:Primary}, in reality, the hydrogen in this inner region should be ionized by \etaA out to a radius of $\sim 120$~AU in Cases~A and B, and everywhere in Case~C. However, the absence of an \etaA ionizing source in our simulations prevents this from occurring. Nonetheless, the absence of an \etaA source in our simulations reveals an important result that may otherwise be missed, namely, that the high optical depth of the inner WWIR prevents any \etaB ionizing photons from penetrating into the inner \etaA wind. The lack of any regions of ionized hydrogen in the unshocked primary wind on the periastron side of the system implies that regardless of the ionization structure of \etaA's innermost wind, ionizing photons from \etaB cannot penetrate the inner WWIR or significantly affect the dense \etaA wind on the periastron side of the system at times around apastron.

\begin{figure*}
  \begin{center}
   \includegraphics[width=174mm]{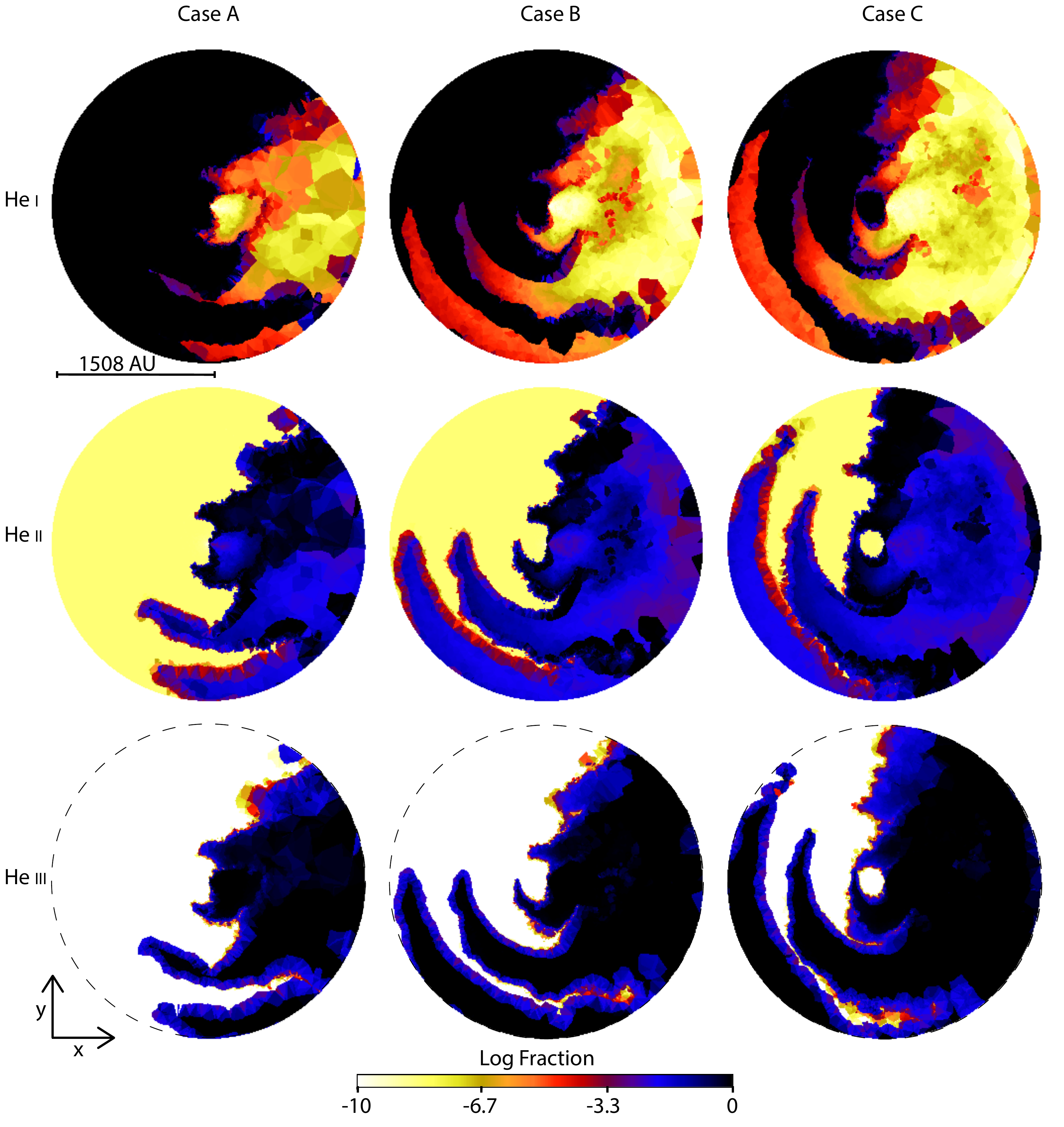}
       \caption{Same as Figure~\ref{fig:dens_ionH_xy}, but with rows showing, from top to bottom, the computed fractions of \HeI, \HeII and \HeIII (log scale). In this and future plots of \HeIII, the dashed circle marks the edge of the spherical computational domain.}\label{fig:ionHe_xy}
  \end{center}
\end{figure*}

Figure~\ref{fig:ionHe_xy} illustrates the fractions of \HeI, \HeII and \HeIII in the orbital plane for the three \MdotA simulations. Comparing to Figure~\ref{fig:dens_ionH_xy}, we see that the regions of \HeIII correlate strongly with the regions of \HII. As expected, the regions of \HI and \HeI are also correlated. The fully-ionized nature of helium in the lower-density \etaB wind is due to the presence of large volumes of very-high-temperature shocked gas, plus the relatively close proximity of such gas to the hot, luminous \etaB ionizing source. As with hydrogen, the helium in the denser primary wind is neutral. The trends as a function of \MdotA seen in Figure~\ref{fig:dens_ionH_xy} for the hydrogen ionization structure are also apparent in the plots of \HeI and \HeIII. This is a key result, as it implies that even with the lower \MdotA of Case~C, \etaB's He-ionizing radiation cannot penetrate significantly the dense WWIRs.

The structure of \HeII (middle row of Figure~\ref{fig:ionHe_xy}) is more involved than that of \HeI and \HeIII. Interestingly, significant fractions of \HeII are principally located in the high-density walls of the WWIRs and outer edges of the dense fingers of \etaA wind that define the low-density fingers of \etaB wind. Regions of lower-temperature unshocked \etaB wind also consist of mostly \HeII. The \HeII is seen primarily as a marker for the transition between the regions of \HeI and \HeIII. For this reason \HeII appears to be an excellent tracer for the high-density compressed post-shock \etaA gas.

\subsubsection{The $xz$ and $yz$ planes}

\begin{figure*}
  \begin{center}
   \includegraphics[width=174mm]{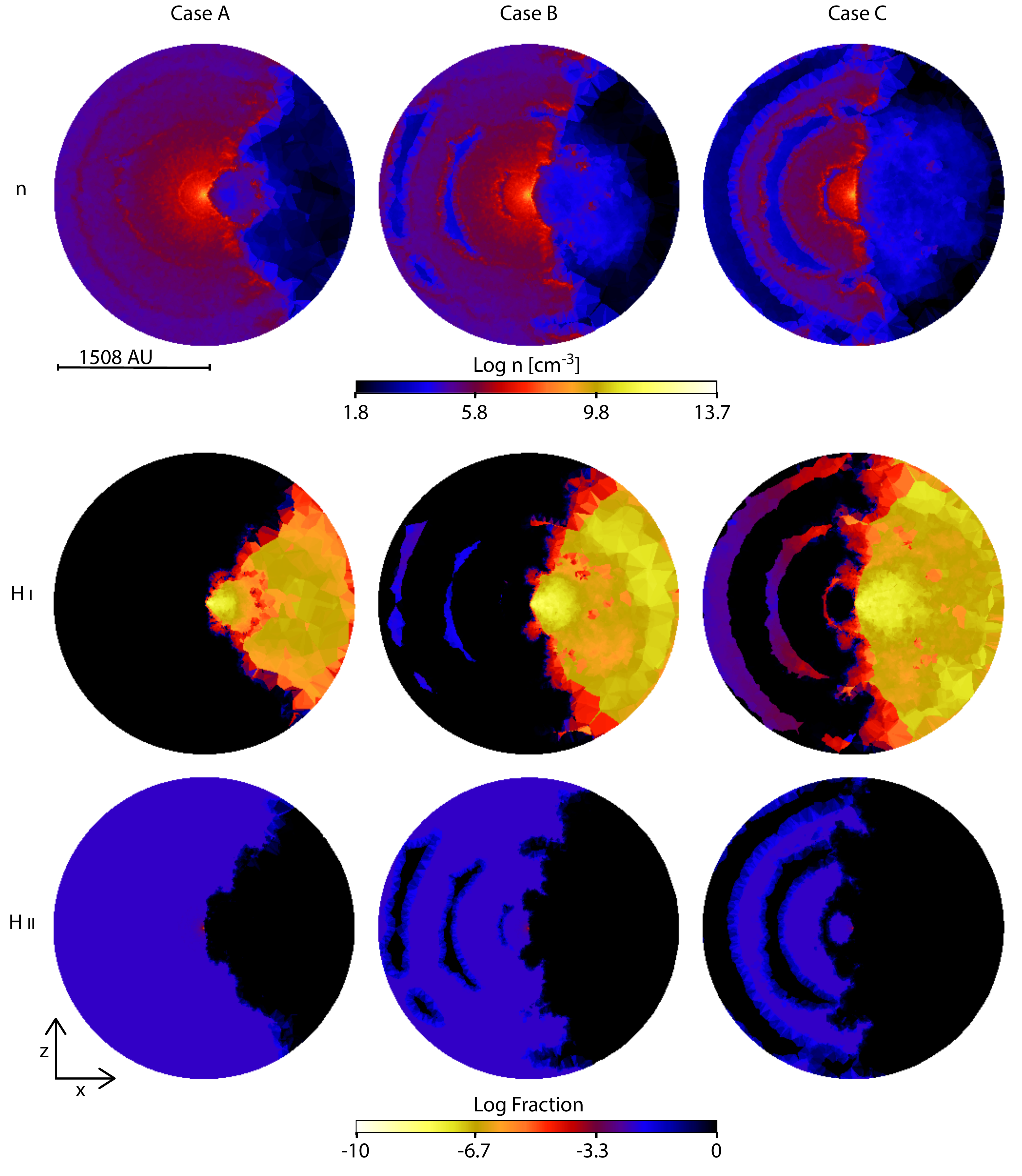}
       \caption{Same as Figure~\ref{fig:dens_ionH_xy}, but for slices centered in the $xz$ plane.}\label{fig:dens_ionH_xz}
  \end{center}
\end{figure*}

\begin{figure*}
  \begin{center}
   \includegraphics[width=174mm]{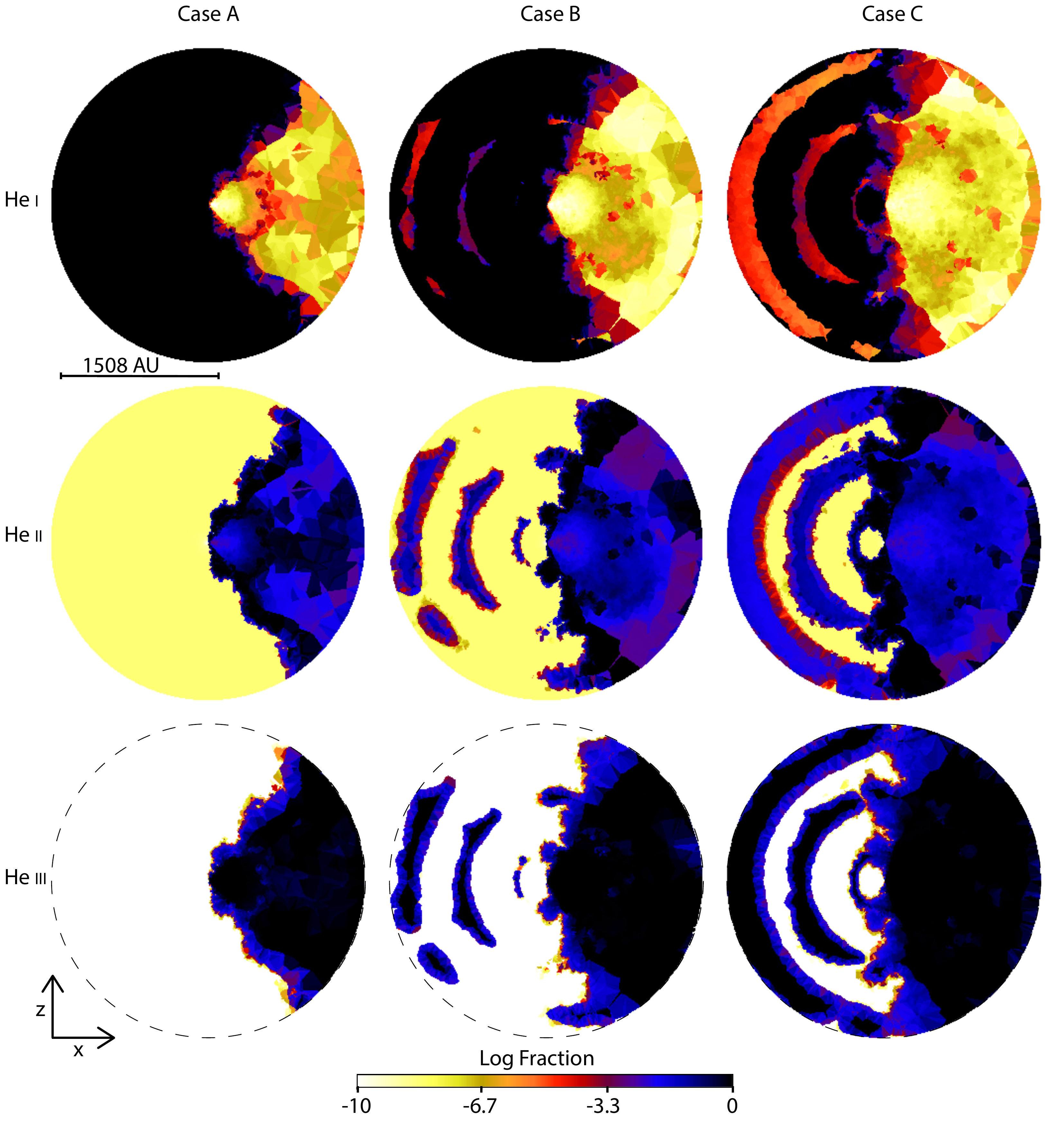}
       \caption{Same as Figure~\ref{fig:ionHe_xy}, but for slices centered in the $xz$ plane.}\label{fig:ionHe_xz}
  \end{center}
\end{figure*}

\begin{figure*}
  \begin{center}
    \includegraphics[width=174mm]{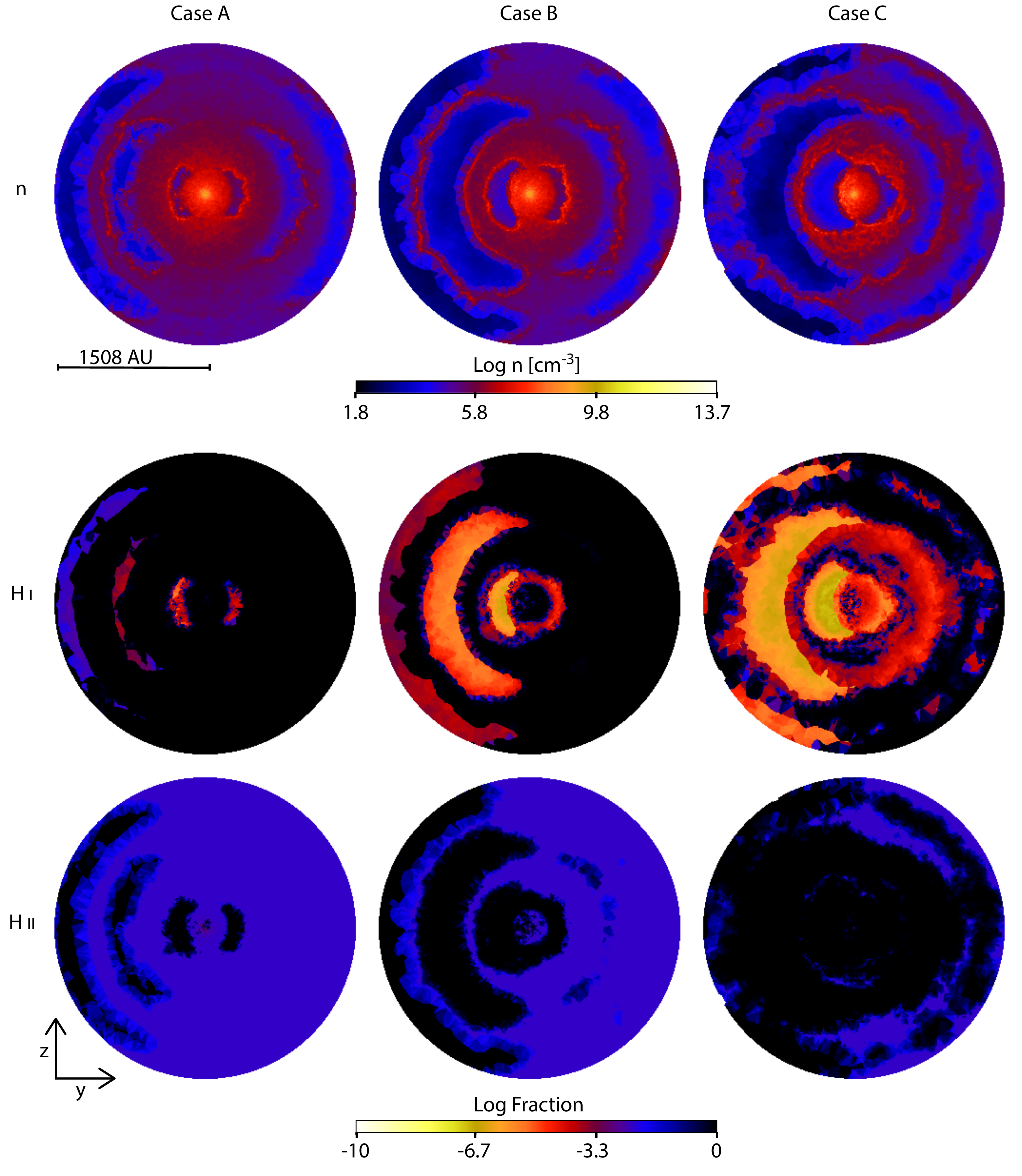}
       \caption{Same as Figure~\ref{fig:dens_ionH_xy}, but for slices centered in the $yz$ plane.}\label{fig:dens_ionH_yz}
  \end{center}
\end{figure*}

\begin{figure*}
  \begin{center}
  \includegraphics[width=174mm]{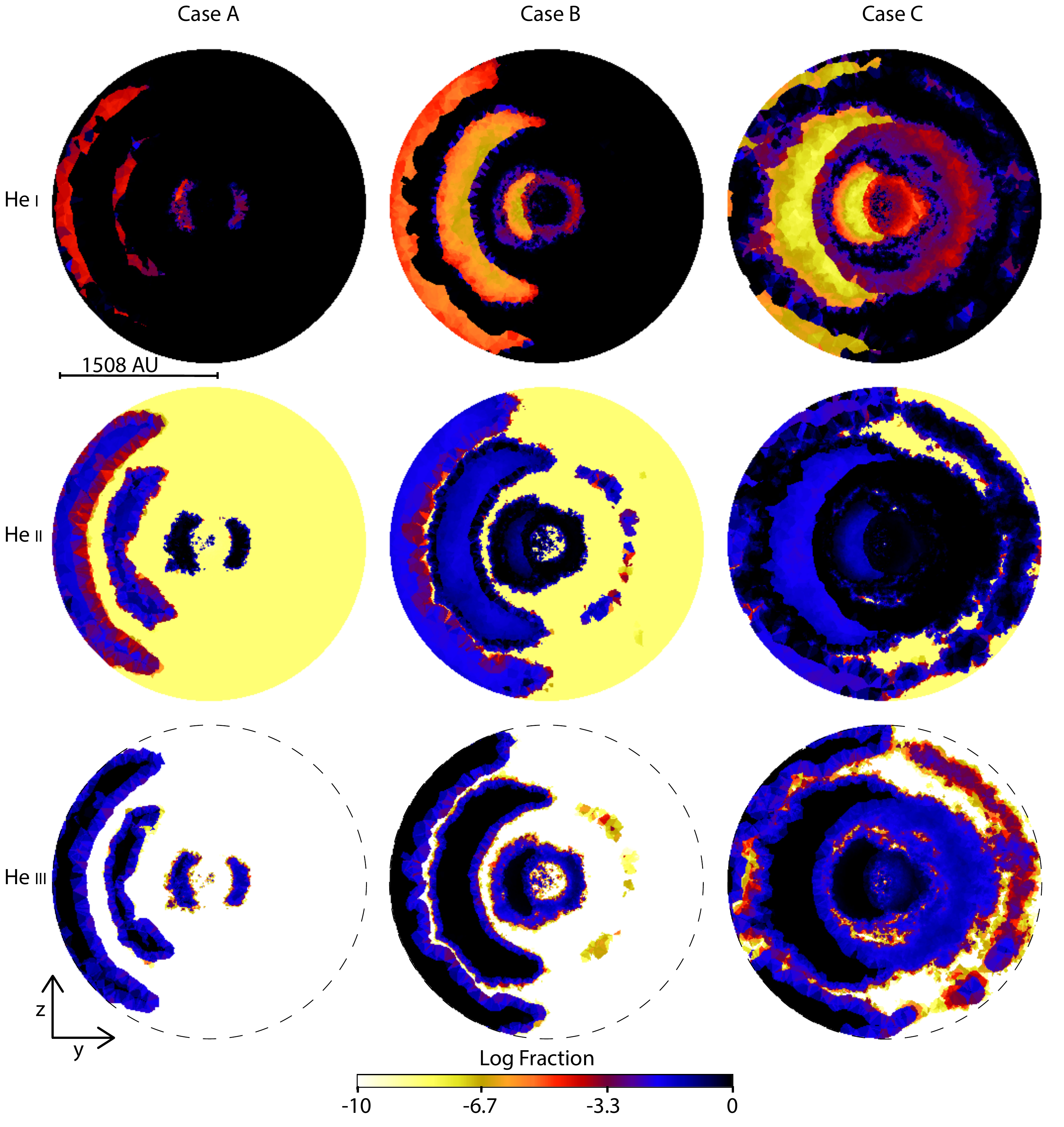}
       \caption{Same as Figure~\ref{fig:ionHe_xy}, but for slices centered in the $yz$ plane.}\label{fig:ionHe_yz}
  \end{center}
\end{figure*}

To help the reader more fully appreciate the complex 3D structure of the simulation results, Figures~\ref{fig:dens_ionH_xz}--\ref{fig:ionHe_yz} present slices showing the number density and H and He ionization structures in the $xz$ and $yz$ planes for each \MdotA. The differences in ionization structure between the three \MdotA are even more apparent in these two planes. There is a clear left-right asymmetry in the density and ionization structure in each panel of the Figures. As in Figures~\ref{fig:dens_ionH_xy} and \ref{fig:ionHe_xy}, the regions of \HeIII correlate strongly with the regions of \HII, while regions of \HeI correlate with those of \HI. The overall volume of ionized material increases with decreasing \MdotA.

Figure~\ref{fig:dens_ionH_xz} shows that the higher the value of \MdotA, the smaller the wind cavities carved by \etaB on the left (periastron) side of the system. They are practically invisible for Case~A. As a result, hydrogen and helium both appear neutral on the left in the Case~A panels. Only the large \etaB wind cavity on the apastron side of the system is ionized in Case~A. Figures~\ref{fig:dens_ionH_xz} and \ref{fig:ionHe_xz} illustrate how the wind cavities on both the periastron and apastron sides of the system are much larger and remain hot for Cases~B and C, resulting in well defined regions of ionized hydrogen and helium.

The top row of Figure~\ref{fig:dens_ionH_xz} also shows clear differences with \MdotA in the density and fragmentation of the dense shell of \etaA wind material on the right (apastron) side of the system. In Case~A, the dense shell is more or less intact, while in Cases~B and C it has fragmented considerably and started to mix with the lower-density \etaB wind. This fragmenting shell produces an interesting hydrogen ionization structure on the right (apastron) side of the system that consists of an inner and outer region of low-density, highly-ionized \etaB wind (in yellow/orange, middle row of Figure~\ref{fig:dens_ionH_xz}) separated by a diffuse shell of denser, less-ionized \etaA wind (in red). The middle row of Figure~\ref{fig:ionHe_xz} again shows that the \HeII is located in the high-density walls of the WWIRs and outer edges of the dense fingers of \etaA wind that define the low-density fingers of \etaB wind, thus tracing the compressed post-shock \etaA gas.

Figure~\ref{fig:dens_ionH_yz} shows that the cavities carved on the right ($+y$) side are always smaller than the ones carved on the left ($-y$), regardless of the value of \MdotA. However, the difference in cavity size between the $+y$ and $-y$ sides increases with decreasing \MdotA. The larger cavities on the left ($-y$) side are also hotter, resulting in well-defined regions of ionized H and He concentrated on the left side. Finally, we see that the shells of dense, compressed \etaA wind on the left are thicker and remain intact longer the higher the \MdotA, reducing the overall volume of ionized material. We note that in reality, H should be ionized everywhere in Case~C.


\section{Discussion}\label{sec:Discussion}

\begin{figure}
  \begin{center}
  \includegraphics[width=80mm]{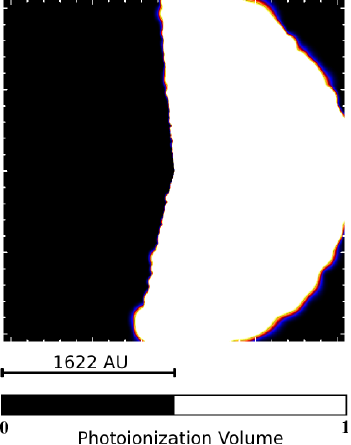}
    \caption{Slice in the $xy$ orbital plane showing the hydrogen photoionization region created by \etaB (white $=$ ionized) for the Case~A simulation at apastron, computed using the approach in \citetalias{Madura_etA_2012}.}\label{fig:M12Model}
  \end{center}
\end{figure}

A major goal of this work was to improve upon the simple approach of \citetalias{Madura_etA_2012} for computing the highly-ionized regions in the \ec system where various observed forbidden emission lines form. The ionization volumes in \citetalias{Madura_etA_2012} were based on geometrical criteria combined with a density threshold, and considered only photoionization of hydrogen due to \etaB. Figure~\ref{fig:M12Model} shows an example of the photoionization region in the orbital plane for Case~A at a phase near apastron, computed using the methods of \citetalias{Madura_etA_2012}. The result is a rather large Str\"{o}mgren-sphere-like volume that predicts the distance that \HI ionizing photons from \etaB can travel. Comparing this to the \SimpleX results in Figures~\ref{fig:dens_ionH_xy} and \ref{fig:ionHe_xy} we clearly see that the \SimpleX method does a significantly improved job at computing the detailed structure of the various ionization volumes, including the penetration of \etaB's photons into the fingers of low-density wind carved within the optically-thick wind of \etaA. The approach of \citetalias{Madura_etA_2012} does not account for the extended WWIR arcs on the apastron side and thus overestimates the ionization extent in these regions.

Effects due to collisional ionization and recombination were also not considered by \citetalias{Madura_etA_2012}. In addition to missing details in the ionization of the low-density fingers of \etaB wind on the apastron side, subtle variations in the ionization of \etaA's wind and the WWIRs on the periastron side, due to recombination, are also absent in the \citetalias{Madura_etA_2012} results. More importantly, the \citetalias{Madura_etA_2012} model does not compute any ionization fractions. The ion fraction is estimated using tables and assuming collisional ionization equilibrium. In contrast, the \SimpleX method computes detailed ionization fractions for both hydrogen and helium. This provides estimates of the extent and magnitude of the ionization as a function of energy, previously unavailable information that is important for determining where the forbidden lines of different ionization potential form. Such information is also crucial for placing constraints on \etaB's ionizing flux.

\citetalias{Madura_etA_2012} found that the observed broad forbidden line emission \citep{Gull_etA_2009, Gull_etA_2011} depends strongly on \MdotA and the ionizing flux from \etaB. \citetalias{Madura_etA_2013} suggested that if the flux from \etaB remains constant, but \MdotA drops by a factor of 2 or more from an initial value of $\approx 8.5 \times 10^{-4} \Msy$, then the photoionization region created by \etaB should increase considerably in size. The results of the \SimpleX simulations in Figures~\ref{fig:dens_ionH_xy}--\ref{fig:ionHe_yz} confirm this, implying that any recent decrease in \MdotA (as speculated by e.g. \citealt{Mehner_etA_2010, Mehner_etA_2011, Mehner_etA_2012}) should greatly change the spatial extent, location, and flux of the observed broad high-ionization forbidden emission lines. The results of this paper will be used in future work to compute synthetic slit-spectral observations of various forbidden lines (e.g. \FeII, \FeIII) for comparison to recent \citep{Gull_etA_2011, Teodoro_etA_2013} and planned observations of \ec from \emph{HST}/STIS. Comparison of the synthetic and observational data can be used to place additional constraints on any recent changes in \MdotA, important for determining \ec's near- and long-term fates \citepalias{Madura_etA_2013}. The improved \SimpleX models will also be useful for refining the orbital orientation parameters obtained by \citetalias{Madura_etA_2012}, and possibly also the stellar wind parameters and/or wind momentum ratio.

\section{Summary and Conclusion}\label{sec:Summary}

We showed that using \SimpleX for the post-processing of 3D SPH simulation output is a viable method to investigate the ionization state of the gas in a complicated colliding wind binary like \ec. \SimpleX provides detailed 3D results of the ionization volumes and fractions for various species of interest, in this case hydrogen and helium, and improves greatly upon simpler approaches such as that in \citetalias{Madura_etA_2012}. Below we summarize our most important results.

\begin{enumerate}[leftmargin=*, label=\arabic*.]

\item The unstructured \SimpleX mesh reproduces everywhere the features present in the original 3D SPH simulation data, leading to a density map that is in excellent agreement with the original SPH one, even where sharp gradients are present. \SimpleX also preserves the high spatial resolution of the original SPH data.\

\item The inclusion of collisional ionization changes the ionization structure of hydrogen and helium most notably in the under-dense fingers of \etaB wind that form between the dense shells of \etaA wind created every periastron passage. Since these regions are typically shielded from \etaB's ionizing flux, including collisional ionization is important to achieve a more complete description of the total ionized volume.

\item Collisional ionization is important in reducing the total optical depth within regions composed of hot \etaB wind that are heated to high temperatures by the various wind-wind collisions. This increases the efficiency of photoionization by \etaB, allowing portions of the dense areas of post-shock \etaA wind and WWIRs on the apastron side of the system to be ionized to varying degrees.

\item The \SimpleX simulations show that the dense, innermost WWIR prevents the \etaB ionizing radiation from penetrating far into the inner wind of \etaA. At phases near apastron, hydrogen and helium ionization are concentrated on the apastron side of the system, with the periastron side consisting of mostly neutral \etaA wind. However, as \MdotA is decreased, low-density fingers of ionized \etaB wind penetrate the dense \etaA wind on the periastron side.

\item We find regions of \HeIII correlate strongly with regions of \HII, while regions of \HI strongly correlate with those of \HeI. \HeII is more complex and primarily marks the transition between the regions of \HeI and \HeIII. \HeII appears to be an excellent tracer for the dense, compressed post-shock \etaA gas and WWIRs.

\item Changing \MdotA results in quite different ionization volumes, with much more ionized gas present for lower \MdotA. Significant variations in ionization structure due to changes in \MdotA are clearly apparent in the $xz$ and $yz$ planes as well as the orbital plane.

\item The large apparent changes in ionization volume with decreasing \MdotA imply that any major decrease in \MdotA should lead to significant observable changes in the spatial extent, location, and flux of the broad high-ionization forbidden emission lines. Future models based on the \SimpleX results may be used to constrain any such potential changes.

\end{enumerate}

In addition to helping us understand \ec's recent mass-loss history, the past \citep{Gull_etA_2011, Teodoro_etA_2013} and future \emph{HST}/STIS spatial maps of \ec's high-ionization forbidden emission lines are a powerful tool that can potentially be used to better determine the nature of the unseen companion star \etaB. Specifically, detailed 3D models of the forbidden line emission based on \SimpleX results like those presented here may allow us to place tighter constraints on \etaB's ionizing flux. This could then be compared to stellar models for a range of O \citep{Martins_etA_2005} and WR \citep{Crowther_2007} stars, providing a more accurate estimate of \etaB's luminosity and temperature.

While applied here to the specific case of \ec, \SimpleX can be used to study numerous other colliding wind binaries or similar systems of astrophysical interest. Application of the \SimpleX algorithm is also not limited to the post-processing of SPH simulation data, output from grid-based codes that use adaptive mesh refinement (AMR) may also be analyzed using \SimpleX \citep{Kruip_2011}. And although \SimpleX has been used in this paper to post-process hydrodynamical simulation data, this work helps set the stage for a future coupling of \SimpleX with the SPH method in order to perform 3D time-dependent radiation-hydrodynamics simulations of complex astrophysical phenomena \citep[see e.g.][]{Pelupessy_etA_2013}.


\section*{Acknowledgments}

T.~I.~M. is supported by an appointment to the NASA Postdoctoral Program at the Goddard Space Flight Center, administered by Oak Ridge Associated Universities through a contract with NASA. Support for T.~R.~G was through programs \#12013, 12508, 12750, 13054, and 13395, provided by NASA through a grant from the Space Telescope Science Institute, which is operated by the Association of Universities for Research in Astronomy, Inc., under NASA contract NAS 5-26555.


\bibliography{biblio}

\begin{thebibliography}{56}
\expandafter\ifx\csname natexlab\endcsname\relax\def\natexlab#1{#1}\fi

\bibitem[{{Corcoran} {et~al}\mbox{.}(2010){Corcoran}, {Hamaguchi}, {Pittard},
  {Russell}, {Owocki}, {Parkin}, \& {Okazaki}}]{Corcoran_etA_2010}
{Corcoran} M.~F., {Hamaguchi} K., {Pittard} J.~M., {Russell} C.~M.~P., {Owocki}
  S.~P., {Parkin} E.~R., {Okazaki} A., 2010, \apj, 725, 1528

\bibitem[{{Corcoran} {et~al}\mbox{.}(2001){Corcoran}, {Ishibashi}, {Swank}, \&
  {Petre}}]{Corcoran_etA_2001}
{Corcoran} M.~F., {Ishibashi} K., {Swank} J.~H., {Petre} R., 2001, \apj, 547,
  1034

\bibitem[{{Crowther}(2007)}]{Crowther_2007}
{Crowther} P.~A., 2007, \araa, 45, 177

\bibitem[{{Damineli}, {Conti} \& {Lopes}(1997){Damineli}, {Conti}, \&
  {Lopes}}]{Damineli_etA_1997}
{Damineli} A., {Conti} P.~S., {Lopes} D.~F., 1997, \na, 2, 107

\bibitem[{{Damineli} {et~al}\mbox{.}(2008{\natexlab{a}}){Damineli}, {Hillier},
  {Corcoran}, {Stahl}, {Groh}, {Arias}, {Teodoro}, {Morrell}, {Gamen},
  {Gonzalez}, {Leister}, {Levato}, {Levenhagen}, {Grosso}, {Colombo}, \&
  {Wallerstein}}]{Damineli_etA_2008_b}
{Damineli} A. {et~al.}, 2008{\natexlab{a}}, \mnras, 386, 2330

\bibitem[{{Damineli} {et~al}\mbox{.}(2008{\natexlab{b}}){Damineli}, {Hillier},
  {Corcoran}, {Stahl}, {Levenhagen}, {Leister}, {Groh}, {Teodoro}, {Albacete
  Colombo}, {Gonzalez}, {Arias}, {Levato}, {Grosso}, {Morrell}, {Gamen},
  {Wallerstein}, \& {Niemela}}]{Damineli_etA_2008_a}
{Damineli} A. {et~al.}, 2008{\natexlab{b}}, \mnras, 384, 1649

\bibitem[{{Davidson} \& {Humphreys}(1997)}]{DavidsonHumphreys_1997}
{Davidson} K., {Humphreys} R.~M., 1997, \araa, 35, 1

\bibitem[{{Fahed} {et~al}\mbox{.}(2011){Fahed}, {Moffat}, {Zorec}, {Eversberg},
  {Chen{\'e}}, {Alves}, {Arnold}, {Bergmann}, {Corcoran}, {Correia Viegas},
  {Dougherty}, {Fernando}, {Fr{\'e}mat}, {Gouveia Carreira}, {Hunger},
  {Knapen}, {Leadbeater}, {Marques Dias}, {Martayan}, {Morel}, {Pittard},
  {Pollock}, {Rauw}, {Reinecke}, {Ribeiro}, {Romeo}, {S{\'a}nchez-Gallego},
  {Dos Santos}, {Schanne}, {Stahl}, {Stober}, {Stober}, {Vollmann}, \&
  {Williams}}]{Fahed_etA_2011}
{Fahed} R. {et~al.}, 2011, \mnras, 418, 2

\bibitem[{{Ferland} {et~al}\mbox{.}(1998){Ferland}, {Korista}, {Verner},
  {Ferguson}, {Kingdon}, \& {Verner}}]{Ferland_etA_1998}
{Ferland} G.~J., {Korista} K.~T., {Verner} D.~A., {Ferguson} J.~W., {Kingdon}
  J.~B., {Verner} E.~M., 1998, \pasp, 110, 761

\bibitem[{{Franco}, {Tenorio-Tagle} \& {Bodenheimer}(1990){Franco},
  {Tenorio-Tagle}, \& {Bodenheimer}}]{Franco_etA_1990}
{Franco} J., {Tenorio-Tagle} G., {Bodenheimer} P., 1990, \apj, 349, 126

\bibitem[{{Gayley}, {Owocki} \& {Cranmer}(1997){Gayley}, {Owocki}, \&
  {Cranmer}}]{Gayley_etA_1997}
{Gayley} K.~G., {Owocki} S.~P., {Cranmer} S.~R., 1997, \apj, 475, 786

\bibitem[{{Groh} {et~al}\mbox{.}(2012){Groh}, {Hillier}, {Madura}, \&
  {Weigelt}}]{Groh_etA_2012}
{Groh} J.~H., {Hillier} D.~J., {Madura} T.~I., {Weigelt} G., 2012, \mnras, 423,
  1623

\bibitem[{{Groh} {et~al}\mbox{.}(2010){Groh}, {Nielsen}, {Damineli}, {Gull},
  {Madura}, {Hillier}, {Teodoro}, {Driebe}, {Weigelt}, {Hartman}, {Kerber},
  {Okazaki}, {Owocki}, {Millour}, {Murakawa}, {Kraus}, {Hofmann}, \&
  {Schertl}}]{Groh_etA_2010}
{Groh} J.~H. {et~al.}, 2010, \aap, 517, A9

\bibitem[{{Gull} {et~al}\mbox{.}(2011){Gull}, {Madura}, {Groh}, \&
  {Corcoran}}]{Gull_etA_2011}
{Gull} T.~R., {Madura} T.~I., {Groh} J.~H., {Corcoran} M.~F., 2011, \apjl, 743,
  L3

\bibitem[{{Gull} {et~al}\mbox{.}(2009){Gull}, {Nielsen}, {Corcoran}, {Madura},
  {Owocki}, {Russell}, {Hillier}, {Hamaguchi}, {Kober}, {Weis}, {Stahl}, \&
  {Okazaki}}]{Gull_etA_2009}
{Gull} T.~R. {et~al.}, 2009, \mnras, 396, 1308

\bibitem[{{Hamaguchi} {et~al}\mbox{.}(2007){Hamaguchi}, {Corcoran}, {Gull},
  {Ishibashi}, {Pittard}, {Hillier}, {Damineli}, {Davidson}, {Nielsen}, \&
  {Kober}}]{Hamaguchi_etA_2007}
{Hamaguchi} K. {et~al.}, 2007, \apj, 663, 522

\bibitem[{{Henley} {et~al}\mbox{.}(2008){Henley}, {Corcoran}, {Pittard},
  {Stevens}, {Hamaguchi}, \& {Gull}}]{Henley_etA_2008}
{Henley} D.~B., {Corcoran} M.~F., {Pittard} J.~M., {Stevens} I.~R., {Hamaguchi}
  K., {Gull} T.~R., 2008, \apj, 680, 705

\bibitem[{{Hillier} {et~al}\mbox{.}(2001){Hillier}, {Davidson}, {Ishibashi}, \&
  {Gull}}]{Hillier_etA_2001}
{Hillier} D.~J., {Davidson} K., {Ishibashi} K., {Gull} T., 2001, \apj, 553, 837

\bibitem[{{Hillier} {et~al}\mbox{.}(2006){Hillier}, {Gull}, {Nielsen},
  {Sonneborn}, {Iping}, {Smith}, {Corcoran}, {Damineli}, {Hamann}, {Martin}, \&
  {Weis}}]{Hillier_etA_2006}
{Hillier} D.~J. {et~al.}, 2006, \apj, 642, 1098

\bibitem[{{Kruip}(2011)}]{Kruip_2011}
{Kruip} C., 2011, PhD thesis, University of Leiden, Leiden, the Netherlands

\bibitem[{{Kruip} {et~al}\mbox{.}(2010){Kruip}, {Paardekooper}, {Clauwens}, \&
  {Icke}}]{Kruip_etA_2010}
{Kruip} C.~J.~H., {Paardekooper} J.-P., {Clauwens} B.~J.~F., {Icke} V., 2010,
  \aap, 515, A78

\bibitem[{{Lef{\`e}vre} {et~al}\mbox{.}(2005){Lef{\`e}vre}, {Marchenko},
  {L{\'e}pine}, {Moffat}, {Acker}, {Harries}, {Annuk}, {Bohlender}, {Demers},
  {Grosdidier}, {Hill}, {Morrison}, {Knauth}, {Skalkowski}, \&
  {Viti}}]{Lefevre_etA_2005}
{Lef{\`e}vre} L. {et~al.}, 2005, \mnras, 360, 141

\bibitem[{{Madura} \& {Groh}(2012)}]{Madura_Groh_2012}
{Madura} T.~I., {Groh} J.~H., 2012, \apjl, 746, L18

\bibitem[{{Madura} {et~al}\mbox{.}(2013){Madura}, {Gull}, {Okazaki}, {Russell},
  {Owocki}, {Groh}, {Corcoran}, {Hamaguchi}, \& {Teodoro}}]{Madura_etA_2013}
{Madura} T.~I. {et~al.}, 2013, \mnras, 436, 3820

\bibitem[{{Madura} {et~al}\mbox{.}(2012){Madura}, {Gull}, {Owocki}, {Groh},
  {Okazaki}, \& {Russell}}]{Madura_etA_2012}
{Madura} T.~I., {Gull} T.~R., {Owocki} S.~P., {Groh} J.~H., {Okazaki} A.~T.,
  {Russell} C.~M.~P., 2012, \mnras, 420, 2064

\bibitem[{{Martins}, {Schaerer} \& {Hillier}(2005){Martins}, {Schaerer}, \&
  {Hillier}}]{Martins_etA_2005}
{Martins} F., {Schaerer} D., {Hillier} D.~J., 2005, \aap, 436, 1049

\bibitem[{{Mehner} {et~al}\mbox{.}(2010){Mehner}, {Davidson}, {Ferland}, \&
  {Humphreys}}]{Mehner_etA_2010}
{Mehner} A., {Davidson} K., {Ferland} G.~J., {Humphreys} R.~M., 2010, \apj,
  710, 729

\bibitem[{{Mehner} {et~al}\mbox{.}(2012){Mehner}, {Davidson}, {Humphreys},
  {Ishibashi}, {Martin}, {Ruiz}, \& {Walter}}]{Mehner_etA_2012}
{Mehner} A., {Davidson} K., {Humphreys} R.~M., {Ishibashi} K., {Martin} J.~C.,
  {Ruiz} M.~T., {Walter} F.~M., 2012, \apj, 751, 73

\bibitem[{{Mehner} {et~al}\mbox{.}(2011){Mehner}, {Davidson}, {Martin},
  {Humphreys}, {Ishibashi}, \& {Ferland}}]{Mehner_etA_2011}
{Mehner} A., {Davidson} K., {Martin} J.~C., {Humphreys} R.~M., {Ishibashi} K.,
  {Ferland} G.~J., 2011, \apj, 740, 80

\bibitem[{{Monnier}, {Tuthill} \& {Danchi}(1999){Monnier}, {Tuthill}, \&
  {Danchi}}]{Monnier_etA_1999}
{Monnier} J.~D., {Tuthill} P.~G., {Danchi} W.~C., 1999, \apjl, 525, L97

\bibitem[{{Okazaki} {et~al}\mbox{.}(2008){Okazaki}, {Owocki}, {Russell}, \&
  {Corcoran}}]{Okazaki_etA_2008}
{Okazaki} A.~T., {Owocki} S.~P., {Russell} C.~M.~P., {Corcoran} M.~F., 2008,
  \mnras, 388, L39

\bibitem[{{Osterbrock} \& {Ferland}(2006)}]{Osterbrock_etA_2006}
{Osterbrock} D.~E., {Ferland} G.~J., 2006, {Astrophysics of gaseous nebulae and
  active galactic nuclei}

\bibitem[{{Owocki}(2007)}]{Owocki_2007}
{Owocki} S., 2007, in Astronomical Society of the Pacific Conference Series,
  Vol. 367, Massive Stars in Interactive Binaries, {St.-Louis} N., {Moffat}
  A.~F.~J., eds., p. 233

\bibitem[{{Paardekooper}, {Kruip} \& {Icke}(2010){Paardekooper}, {Kruip}, \&
  {Icke}}]{Paardekooper_etA_2010}
{Paardekooper} J.-P., {Kruip} C.~J.~H., {Icke} V., 2010, \aap, 515, A79

\bibitem[{{Paardekooper} {et~al}\mbox{.}(2011){Paardekooper}, {Pelupessy},
  {Altay}, \& {Kruip}}]{Paardekooper_etA_2011}
{Paardekooper} J.-P., {Pelupessy} F.~I., {Altay} G., {Kruip} C.~J.~H., 2011,
  \aap, 530, A87

\bibitem[{{Parkin} {et~al}\mbox{.}(2011){Parkin}, {Pittard}, {Corcoran}, \&
  {Hamaguchi}}]{Parkin_etA_2011}
{Parkin} E.~R., {Pittard} J.~M., {Corcoran} M.~F., {Hamaguchi} K., 2011, \apj,
  726, 105

\bibitem[{{Parkin} {et~al}\mbox{.}(2009){Parkin}, {Pittard}, {Corcoran},
  {Hamaguchi}, \& {Stevens}}]{Parkin_etA_2009}
{Parkin} E.~R., {Pittard} J.~M., {Corcoran} M.~F., {Hamaguchi} K., {Stevens}
  I.~R., 2009, \mnras, 394, 1758

\bibitem[{{Parkin} \& {Sim}(2013)}]{Parkin_etA_2013}
{Parkin} E.~R., {Sim} S.~A., 2013, \apj, 767, 114

\bibitem[{{Pawlik} \& {Schaye}(2008)}]{Pawlik_etA_2008}
{Pawlik} A.~H., {Schaye} J., 2008, \mnras, 389, 651

\bibitem[{{Pelupessy} {et~al}\mbox{.}(2013){Pelupessy}, {van Elteren}, {de
  Vries}, {McMillan}, {Drost}, \& {Portegies Zwart}}]{Pelupessy_etA_2013}
{Pelupessy} F.~I., {van Elteren} A., {de Vries} N., {McMillan} S.~L.~W.,
  {Drost} N., {Portegies Zwart} S.~F., 2013, \aap, 557, A84

\bibitem[{{Pittard} \& {Corcoran}(2002)}]{Pittard_etA_2002}
{Pittard} J.~M., {Corcoran} M.~F., 2002, \aap, 383, 636

\bibitem[{{Pittard} {et~al}\mbox{.}(1998){Pittard}, {Stevens}, {Corcoran}, \&
  {Ishibashi}}]{Pittard_etA_1998}
{Pittard} J.~M., {Stevens} I.~R., {Corcoran} M.~F., {Ishibashi} K., 1998,
  \mnras, 299, L5

\bibitem[{{Ritzerveld} \& {Icke}(2006)}]{Ritzerveld_etA_2006}
{Ritzerveld} J., {Icke} V., 2006, \pre, 74, 026704

\bibitem[{{Ritzerveld}(2007)}]{Ritzerveld_2007}
{Ritzerveld} N.~G.~H., 2007, PhD thesis, Leiden Observatory, Leiden University,
  P.O.~Box 9513, 2300 RA Leiden, The Netherlands

\bibitem[{Russell(2013)}]{Russell_2013}
Russell C. M.~P., 2013, PhD thesis, University of Delaware, Newark, DE, USA

\bibitem[{{Shapiro} {et~al}\mbox{.}(2006){Shapiro}, {Iliev}, {Alvarez}, \&
  {Scannapieco}}]{Shapiro_etA_2006}
{Shapiro} P.~R., {Iliev} I.~T., {Alvarez} M.~A., {Scannapieco} E., 2006, \apj,
  648, 922

\bibitem[{{Smith}(2006)}]{Smith_2006}
{Smith} N., 2006, \apj, 644, 1151

\bibitem[{{Stevens} \& {Pollock}(1994)}]{Stevens_etA_1994}
{Stevens} I.~R., {Pollock} A.~M.~T., 1994, \mnras, 269, 226

\bibitem[{{Teodoro} {et~al}\mbox{.}(2012){Teodoro}, {Damineli}, {Arias}, {de
  Ara{\'u}jo}, {Barb{\'a}}, {Corcoran}, {Borges Fernandes},
  {Fern{\'a}ndez-Laj{\'u}s}, {Fraga}, {Gamen}, {Gonz{\'a}lez}, {Groh},
  {Marshall}, {McGregor}, {Morrell}, {Nicholls}, {Parkin}, {Pereira},
  {Phillips}, {Solivella}, {Steiner}, {Stritzinger}, {Thompson}, {Torres},
  {Torres}, \& {Zevallos Herencia}}]{Teodoro_etA_2012}
{Teodoro} M. {et~al.}, 2012, \apj, 746, 73

\bibitem[{{Teodoro} {et~al}\mbox{.}(2008){Teodoro}, {Damineli}, {Sharp},
  {Groh}, \& {Barbosa}}]{Teodoro_etA_2008}
{Teodoro} M., {Damineli} A., {Sharp} R.~G., {Groh} J.~H., {Barbosa} C.~L.,
  2008, Monthly Notices of the Royal Astronomical Society, 387, 564

\bibitem[{{Teodoro} {et~al}\mbox{.}(2013){Teodoro}, {Madura}, {Gull},
  {Corcoran}, \& {Hamaguchi}}]{Teodoro_etA_2013}
{Teodoro} M., {Madura} T.~I., {Gull} T.~R., {Corcoran} M.~F., {Hamaguchi} K.,
  2013, \apjl, 773, L16

\bibitem[{{Townsend}(2009)}]{Townsend_2009}
{Townsend} R.~H.~D., 2009, \apjs, 181, 391

\bibitem[{{Tuthill}, {Monnier} \& {Danchi}(1999){Tuthill}, {Monnier}, \&
  {Danchi}}]{Tuthill_etA_1999}
{Tuthill} P.~G., {Monnier} J.~D., {Danchi} W.~C., 1999, \nat, 398, 487

\bibitem[{{Verner}, {Bruhweiler} \& {Gull}(2005){Verner}, {Bruhweiler}, \&
  {Gull}}]{Verner_etA_2005}
{Verner} E., {Bruhweiler} F., {Gull} T., 2005, \apj, 624, 973

\bibitem[{{Whitelock} {et~al}\mbox{.}(2004){Whitelock}, {Feast}, {Marang}, \&
  {Breedt}}]{Whitelock_etA_2004}
{Whitelock} P.~A., {Feast} M.~W., {Marang} F., {Breedt} E., 2004, \mnras, 352,
  447

\bibitem[{{Williams}, {Rauw} \& {van der Hucht}(2009){Williams}, {Rauw}, \&
  {van der Hucht}}]{Williams_etA_2009}
{Williams} P.~M., {Rauw} G., {van der Hucht} K.~A., 2009, \mnras, 395, 2221

\end{thebibliography}

\label{lastpage}

\end{document}